\documentclass[twocolumn, 10pt]{IEEEtran}
\usepackage{kotex}
\usepackage{arydshln}
\usepackage{multirow}

\usepackage{url}
\usepackage[breaklinks]{hyperref} 
\usepackage{breakurl}
\usepackage{graphicx}
\usepackage{amssymb}
\usepackage{amsmath}
\usepackage{mathtools}
\usepackage{cite}
\usepackage{stfloats}
\usepackage{epstopdf}
\usepackage{psfrag}
\usepackage[mathscr]{euscript}
\usepackage{acronym}  
\usepackage{booktabs}
\usepackage{mdframed}
\usepackage[table]{xcolor}
\usepackage{multirow}
\usepackage{rotating}
\usepackage{verbatim}
\usepackage[skip=0.333\baselineskip]{caption}
\usepackage{dcolumn}
\usepackage{amsfonts}
\usepackage{makecell}
\usepackage{array}
\usepackage{hhline}
\usepackage{ctable}
\usepackage{caption}
\usepackage{subcaption}
\usepackage{xparse}
\usepackage{bigints}

\newtheorem{definition}{Definition}
\newtheorem{theorem}{Theorem}

\makeatletter
\def\underbracex#1#2{\mathop{\vtop{\m@th\ialign{##\crcr
				$\hfil\displaystyle{#2}\hfil$\crcr
				\noalign{\kern3\p@\nointerlineskip}%
				#1\crcr\noalign{\kern3\p@}}}}\limits}

\def\upbracefilla{$\m@th \setbox\z@\hbox{$\braceld$}%
	\bracelu\leaders\vrule \@height\ht\z@ \@depth\z@\hfill 
	\kern\p@\vrule \@width\p@\kern\p@\vrule \@width\p@\kern\p@\vrule \@width\p@
	$}

\def\upbracefillb{$\m@th \setbox\z@\hbox{$\braceld$}%
	\vrule \@width\p@\kern\p@\vrule \@width\p@\kern\p@\vrule \@width\p@\kern\p@
	\leaders\vrule \@height\ht\z@ \@depth\z@\hfill\bracerd
	\braceld\leaders\vrule \@height\ht\z@ \@depth\z@\hfill
	\kern\p@\vrule \@width\p@\kern\p@\vrule \@width\p@\kern\p@\vrule \@width\p@
	$}

\def\upbracefillc{$\m@th \setbox\z@\hbox{$\braceld$}%
	\vrule \@width\p@\kern\p@\vrule \@width\p@\kern\p@\vrule \@width\p@\kern\p@
	\leaders\vrule \@height\ht\z@ \@depth\z@\hfill
	\kern\p@\vrule \@width\p@\kern\p@\vrule \@width\p@\kern\p@\vrule \@width\p@
	$}

\def\upbracefilld{$\m@th \setbox\z@\hbox{$\braceld$}%
	\vrule \@width\p@\kern\p@\vrule \@width\p@\kern\p@\vrule \@width\p@\kern\p@
	\leaders\vrule \@height\ht\z@ \@depth\z@\hfill\braceru$}

\def\upbracefillbd{$\m@th \setbox\z@\hbox{$\braceld$}%
	\vrule \@width\p@\kern\p@\vrule \@width\p@\kern\p@\vrule \@width\p@\kern\p@
	\bracerd\braceld
	\leaders\vrule \@height\ht\z@ \@depth\z@\hfill\braceru$}

\makeatother
\acrodef{ITS}{intelligent transportation systems}
\acrodef{C-ITS}{cooperative intelligent transportation systems}
\acrodef{ECU}{electronic control unit}
\acrodef{V2X}{vehicle-to-everything}
\acrodef{IoT}{Internet of Things}
\acrodef{OBU}{on-board unit}
\acrodef{ABE}{attribute-based encryption}
\acrodef{DSRC}{dedicated short-range communications}
\acrodef{WAVE}{wireless access in vehicular environments}
\acrodef{OEM}{original equipment manufacturer}
\acrodef{RSU}{road side unit}
\acrodef{UE}{user equipment}
\acrodef{gNB}{next generation Node B}
\acrodef{CAV}{connected and autonomous vehicle}
\acrodef{ITS}{intelligent transportation system}
\acrodef{IoV}{Internet of Vehicles}
\acrodef{CRT}{Chinese remainder theorem}
\acrodef{SA}{security agent}
\acrodef{AKA}{authentication and key management}
\acrodef{VANET}{vehicular ad hoc network}
\acrodef{JCA}{Java cryptography architecture}
\acrodef{JCE}{Java cryptography extension}
\acrodef{UDM}{unified data management}
\acrodef{ARPF}{authentication credential repository and processing function}
\acrodef{ABKE}{attribute-based key exchange}
\acrodef{CAD}{cooperative autonomous driving}
\acrodef{DDH}{decisional Diffie-Hellman}
\acrodef{5G-V2X}{5G-based V2X}
\acrodef{CP-ABE}{ciphertext-policy ABE}
\acrodef{IIoT}{industrial Internet of Things}
\acrodef{OEEP-ABE}{outsourcing-enabled and enhanced privacy-preserving ABE}
\acrodef{OE-IBS}{outsourcing-enabled IBS}
\acrodef{PS-E2EID}{practical and secure vehicular communication protocol for E2E security to in-vehicle end-devices}
\acrodef{CV}{connected vehicle}
\acrodef{AV}{autonomous vehicle}
\acrodef{ADAS}{advanced driver-assistance system}
\acrodef{E2E}{end-to-end}
\acrodef{SoC}{system-on-chip}
\acrodef{C-IND-CPA-RUCA}{ciphertext indistinguishability against chosen plaintext attack and restricted user collusion attack}
\acrodef{P-IND-CPA-RUCA}{policy indistinguishability against chosen plaintext attack and restricted user collusion attack}
\acrodef{IND-CPA}{indistinguishability against chosen plaintext attack}
\acrodef{PPT}{probabilistic polynomial time}
\acrodef{TPM}{trusted platform module}
\acrodef{C-V2X}{cellular V2X}

\newcommand{\blue}[1]{{\textcolor[rgb]{0,0,1}{#1}}}

\begin{document}

	\newcommand{\paperTitle}{Ensuring End-to-End Security with \\ Fine-grained Access Control for \\ Connected and Autonomous Vehicles}
	
	
	
	\title{\paperTitle}
	
	\author{
		Donghyun Yu, Sungho Lee, Ruei-Hau Hsu, and Jemin Lee
		\vspace{-6mm}
		\thanks{
			Corresponding author is J. Lee.

			D.\ Yu is with the Department of Electrical Engineering and Computer Science, Daegu Gyeongbuk Institute of Science and Technology (DGIST), Daegu 42988, South Korea
			(e-mail: \texttt{xaos4715@dgist.ac.kr}).
			
			S. Lee and J. Lee 
			are with the School of Electrical and Electronic Engineering, Yonsei University, Seoul 03722, South Korea
			(e-mail: \texttt{\{sh.lee23, jemin.lee\}@yonsei.ac.kr}).
			
			R.\ -H.\ Hsu is with the Department of Computer Science and Engineering, National Sun Yat-sen University, Kaohsiung 80424, Taiwan, R.O.C.
			(e-mail: \texttt{rhhsu@mail.cse.nsysu.edu.tw}).	
		}
	}
	
	\maketitle 
	
	%
	
	%
	
	%
	\setcounter{page}{1}
	\acresetall
	
	\begin{abstract}
		As advanced V2X applications emerge in the connected and autonomous vehicle (CAV), 
		the data communications between in-vehicle end-devices and outside nodes increase,
		which make the end-to-end (E2E) security to in-vehicle end-devices as the urgent issue to be handled. 
		However, the E2E security with fine-grained access control still remains as a challenging issue for resource-constrained end-devices since the existing security solutions require complicated key management and high resource consumption. 
		Therefore, we propose a practical and secure vehicular communication protocol for the E2E security based on a new attribute-based encryption (ABE) scheme. In our scheme, the outsourced computation is provided for encryption, and the computation cost for decryption constantly remains small, regardless of the number of attributes. 
		The policy privacy can be ensured by the proposed ABE to support privacy-sensitive V2X applications,
		and the existing identity-based signature for outsourced signing is newly reconstructed. 
		Our scheme achieves the confidentiality, message authentication, identity anonymity, unlinkability, traceability, and reconfigurable outsourced computation, and we also show the practical feasibility of our protocol via the performance evaluation.
	\end{abstract} 
	
	\vspace{-6mm}
	\section{Introduction}\label{sec:Introduction}
	The \acp{CAV} are emerging vehicles based on the convergence technology that combines \acp{CV} and \acp{AV} \cite{ZhaZhaFieYan:18}. For user convenience and safety, \acp{CAV} have been equipped with various sensors such as LiDAR, radar, and cameras. For autonomous driving and vehicular services, the data of surrounding environments is generally collected by those sensors at vehicles, and transmitted to inner nodes such as \acp{ECU} or outside nodes such as other vehicles, \acp{RSU}, \acp{UE}, and \acp{OEM} \cite{MSil:18}.
	The data transmission can be done by vehicular communications that include \emph{in-vehicle communications} for inner nodes and \emph{\ac{V2X} communications} for outside nodes. 
	
	For reliable operations of \acp{CAV}, the vehicular communications need to satisfy certain requirements including low latency, high reliability, scalability, and security.	Among those, the security is the one of the most crucial and important ones	since increasing number of sensors and frequent communications of \acp{CAV} result in wider attack surfaces\cite{PjSse:15}. 
	For example, attackers can have more opportunities to analyze vehicle data for extracting private data such as vehicle movement patterns and locations. 
	The attacks on control data can be even more fatal as the \acp{CAV} use them for autonomous driving and the manipulated control data menaces human safety. 	
	
	For the security of vehicular communications, many works have been presented \cite{WpCcmKs:20, ShaLinLuZuo:16, VijAzeKanDeb:16}. The hybrid device-to-device (D2D) message authentication scheme for source authentication, integrity, conditional privacy, and traceability has been proposed for \ac{VANET}\cite{WpCcmKs:20}.
	For the security of massive \ac{V2X} communications, the group signature scheme has been proposed for threshold authentication, traceability, and revocation\cite{ShaLinLuZuo:16}.
	For the secure \acp{VANET}, the dual key management with two-factor authentication techniques has also been proposed\cite{VijAzeKanDeb:16}. However, although those works provide a higher level of security, they do not consider the \emph{fine-grained access control}, which becomes an essential requirement for efficient and secure data sharing, especially in large vehicular networks.
		
	The fine-grained access control is to grant or deny the data access based on multiple conditions (e.g., attributes or attribute-based policies) for avoiding the data exposure to unauthorized access in the networks. 
	Recently, some works have been presented to provide the fine-grained access control for vehicular communications \cite{KanLiuYaoWanLi:16, GouHarAliGue:21}, generally by using the \ac{ABE} \cite{JbAsBw:07}.
	Specifically, the \ac{CP-ABE} has been used with the identity-based signature for the privacy-preserving message provision in \acp{VANET}\cite{KanLiuYaoWanLi:16}. The \ac{CP-ABE} has also been used for secure and efficient key distribution to support the vehicular cloud computing \cite{GouHarAliGue:21}. 
	However, although the fine-grained access control is provided in those works, the inevitable cost of \ac{ABE}, i.e., \emph{the high computation complexity}, has not been solved, which can result in long processing delay, especially at resource-constrained devices such as \acp{ECU}.

	The high computation cost has been an important issue for the security of resource-constrained devices 
	including sensors and IoT devices. Recently, to solve this issue for vehicle networks, the online/offline encryption and the outsourced decryption have been introduced \cite{FenYuAloAlaLvMum:20, YcJpZl:22, TiaLiQuaChaBak:21}.	Specifically, the outsourced decryption with a parallel computation method has been proposed for edge \ac{IoV}\cite{FenYuAloAlaLvMum:20}. The functional encryption with online/offline decryption has been presented for the decentralized and partial privacy data access control with lightweight decryption in VANET\cite{YcJpZl:22}. The lightweight \ac{ABE} scheme based on the online/offline encryption and the outsourced decryption has also been proposed for data sharing in an \ac{ITS}.
	
	Despite those efforts, the computation costs for encryption and decryption in those works 
	are still high for \acp{ECU}, which generally have limited computing capabilities. Furthermore, the policy privacy is not guaranteed in \cite{FenYuAloAlaLvMum:20} and \cite{YcJpZl:22}, so the policies of data can be exposed to attackers, who may use them to analyze the attributes of certain receivers. In addition, the most existing works including \cite{WpCcmKs:20, ShaLinLuZuo:16, VijAzeKanDeb:16, KanLiuYaoWanLi:16, GouHarAliGue:21, FenYuAloAlaLvMum:20, YcJpZl:22, TiaLiQuaChaBak:21} only focus on the communication security to the \ac{OBU}, and fail to provide the \textit{\ac{E2E} security} to in-vehicle end-devices, such as \acp{ECU} and \ac{ADAS}. In reality,  the \acp{OBU} are relatively more vulnerable to attacks as they are connected to the external vehicular networks, and \acp{ECU} and \acp{ADAS} are the end-devices, which actually transfer and use the driving-related data.
	Therefore, it is essential to prevent data exposure even against unintended in-vehicle nodes to achieve a higher security level. A novel security solution to provide the \textit{\ac{E2E} security} to in-vehicle end-devices is truly demanded for \acp{CAV}. 
	
	Therefore, this paper develops a \textit{\ac{PS-E2EID}} to provide secure and efficient data sharing with fine-grained access control. Our \ac{PS-E2EID} mainly aims to provide not only authentication between a \ac{CAV} and another \ac{CAV}, \ac{RSU}, \ac{UE}, and \ac{OEM}, but also attribute-based data sharing (ABDS) between those and in-vehicle end-devices. For our \ac{PS-E2EID}, by enhancing the proposed \ac{ABE} in \cite{KapTsaSmi:07}, we propose an \ac{OEEP-ABE} scheme as the underlying cryptosystem for fine-grained access control. Our proposed \ac{OEEP-ABE} reduces the computation cost at in-vehicle end-devices depending on lightweight decryption and outsourced techniques for encryption. Additionally, we propose \ac{OE-IBS} with outsourced technology to existing identity-based signature schemes \cite{LjkBjZj:10} to further reduce the computation cost of message signing at in-vehicle end devices.
	The experimental results also show that the execution time of the \ac{PS-E2EID} is sufficiently low. The contributions of this paper are listed as follows: 
	
	\begin{itemize}
		\item We propose \ac{OEEP-ABE} to reduce the computational burden from complex security operations on resource-constrained \acp{ECU} using outsourced encryption and lightweight decryption. Our proposed \ac{OEEP-ABE} achieves attribute-based access control as well as policy privacy against internal attackers. In addition, we propose OE-IBS to reduce the computational burden from ECU in the signature generation process as well.
		\item Using \ac{OEEP-ABE} and \ac{OE-IBS}, we propose the \ac{PS-E2EID} to achieve attribute-based access control, policy privacy, and \textit{\ac{E2E} security} against internal attackers. In addition, the \ac{PS-E2EID} can ensure confidentiality, authentication, identity anonymity, unlinkability, and traceability which have also been achieved in most of the previous V2X security solutions. To the best of our knowledge, this is the first work that achieves the \textit{\ac{E2E} security} to in-vehicle end-devices (e.g., \acp{ECU} and \ac{ADAS}) for secure vehicular communications. 
		\item We show the feasibility of the \ac{PS-E2EID}. Specifically, we measure the execution time of the \ac{PS-E2EID} with different numbers of sending \acp{ECU} as well as those of system and receiver attributes in CANoe \cite{Vector}. The captured execution time shows that the \ac{PS-E2EID} can ensure sufficiently low latency for more secure and faster data sharing.
	\end{itemize}
	\vspace{-2mm}
	\section{System and Security Models}\label{sec:SystemAndSecurityModels}
	This section describes our \ac{C-V2X} network model for the \acp{CAV}. We use 5G, which is the latest cellular network, to present an example of our system model as the 5G-based \ac{C-V2X}.\footnote{3GPP TS 23.287, Available: \url{https://portal.3gpp.org/desktopmodules/Specifications/SpecificationDetails.aspx?specificationId=3578}} We then explain the attack model and security requirements. Finally, we introduce the security model and definitions, which are used to analyze the security of the proposed cryptosystems and protocols.
	\vspace{-2mm}
	\subsection{System Model}\label{subsec:SystemModel}
			\begin{figure}[t!]
		\vspace{-2mm}
		\begin{center}   
			{ 
				\includegraphics[width=0.874\columnwidth]{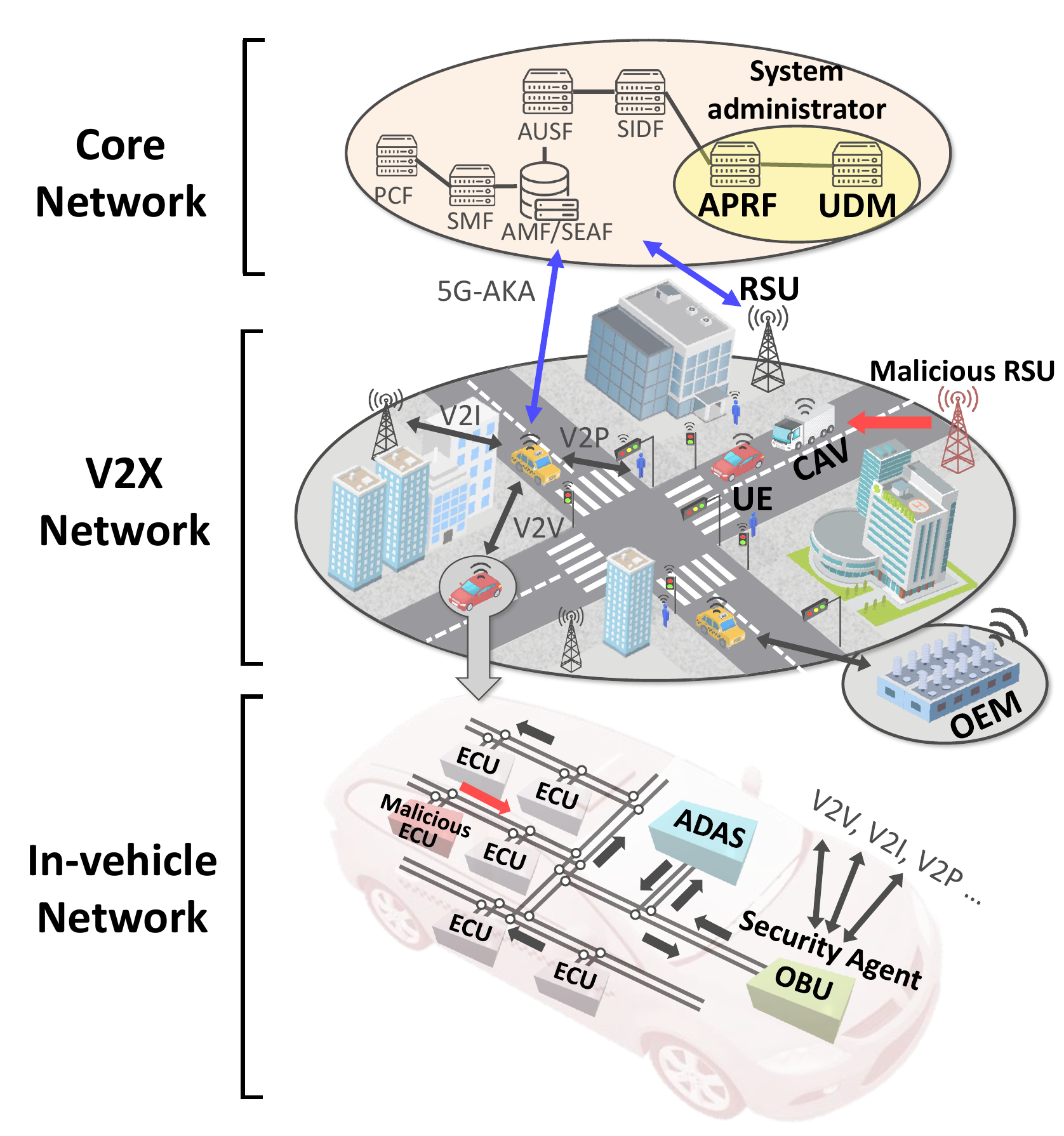}
			}
		\end{center}\vspace{-4mm}
		\caption{
			Example of the system model in 5G.
		}
		\label{fig:SystemModel}
		\vspace{-4mm}
	\end{figure}
	The system model is shown in Fig. \ref{fig:SystemModel}, which is divided into an in-vehicle network, a \ac{V2X} network, and a core network. Each component of C-V2X networks is explained below.
	
		\subsubsection{Core Network Node} The core network has several functions including \ac{UDM} and \ac{ARPF}. They are responsible for registering system members such as \acp{CAV}, \acp{RSU}, \acp{UE}, and \acp{OEM} to the \ac{C-V2X} network and authenticating between members. We consider only \ac{UDM} and \ac{ARPF} as participants in the proposed protocol.
		\begin{itemize}
		\item \ac{UDM}/\ac{ARPF}: The \ac{UDM}/\ac{ARPF} serve as system administrators for C-V2X in the 5G core network. They are regarded as a fully trusted authority with powerful computation and storage resources. They can trace the real identities of system members. Moreover, they can generate, distribute, and manage global public and private parameters of the system members.
		\end{itemize}
		\subsubsection{V2X Network Node} The V2X network nodes such as \acp{CAV}, \acp{RSU}, \acp{UE}, and \acp{OEM} can be both vehicular data receivers and data senders for smooth vehicle driving. 
		\begin{itemize}
		\item \ac{RSU}: The \ac{RSU} collects and processes road traffic data and \ac{CAV} data to generate V2X service data. 
		
		\item \ac{UE}: The \ac{UE} enhances the safety for \acp{CAV} and pedestrians through data sharing with \acp{CAV}.
		
		\item \ac{OEM}: The \ac{OEM} not only checks \acp{CAV} by collecting their status data from \acp{ECU}, but also sends firmware updates for \acp{ECU}. Moreover, it support fast encryption by providing encryption materials to manufactured \acp{ECU}.
		\end{itemize}
		\subsubsection{In-vehicle Network Node} The in-vehicle nodes include an \ac{OBU}, an \ac{ADAS}, and several \acp{ECU} in a \ac{CAV}. They communicate with each other to control various functions.
		\begin{itemize}
		\item \ac{OBU}: The \ac{OBU} is a communication device to connect the \ac{CAV} with the V2X network nodes. It can also supports the computation tasks, requested from \acp{ECU}. We define this type of \ac{OBU} as a \ac{SA}.
		
		\item \ac{ADAS}: The \ac{ADAS} is an intelligent system to assist the driver to make decisions through vehicular communications \cite{Kal:16}.\footnote{Specifically, the \ac{ADAS} refers to \ac{SoC} for the \ac{ADAS} \cite{LLiu:21} (e.g., TDA3x, EyeQ).} It can also send some commands to \acp{ECU} for \ac{CAV} control.
		
		\item \ac{ECU}: The \acp{ECU} generally have low computing power compared to the other system members, and hence require long latency to execute advanced cryptosystems (e.g., public key cryptography). Each \ac{ECU} is equipped with a \ac{TPM} for hardware-based security \cite{TPM}. In our system, the \ac{TPM} securely generates a new signing key according to \ac{ECU}'s pseudo-identity.
		\end{itemize}

Since we consider 5G for our \ac{C-V2X} network model, each system member has already shared a long-term key with UDM/ARPF through the 5G-\ac{AKA}. We consider that all communication channels, used for long-term key sharing, are secure. On the other hand, the other channels such as for vehicular communications can be vulnerable to various attacks.
	\vspace{-2mm}
	\subsection{Attack Model and Security Requirements}\label{subsec:AttackModel}
	In our system model, the \ac{OBU} is the only entity, which can learn sensitive data of \acp{ECU} due to the outsourcing procedure. Hence, we consider the \ac{OBU} as an honest-but-curious adversary that follows the protocol honestly but may attempt to expose all the secret keys and message contents to maximize the advantages of its corruption.
	Next, all system members, except for the OBU (i.e., \acp{ADAS}, \acp{ECU}, \acp{RSU}, \acp{UE}, and \acp{OEM}) have their own assigned attributes. Additionally, they can encrypt messages according to specified policies, and then send the messages to their intended receivers. Any corrupted entity is considered as a malicious one, but it is assumed not to collude with the \ac{OBU} (acting as the \ac{SA}) and the \ac{OEM}. The \ac{OBU} and \ac{OEM}, which receive outsourcing requests for encryption from \acp{ECU}, cannot collude with each other. To capture their abilities, we organize the following security requirements.

	\begin{itemize}
		\item \emph{Confidentiality}:
		It ensures that messages with private data, broadcasted in V2X and in-vehicle networks, are concealed against all attackers.
		\item \emph{Message Authentication}:
		It guarantees that all legal entities in the system can verify the sender of messages and the integrity of received messages.
		\item \emph{Identity Anonymity and Unlinkability}:
		The footprint of a CAV can be tracked by attackers using V2X services. Hence, to ensure anonymity, it is important to prevent attackers from identifying whether authentication messages from any two sessions are sent by the same sender or not.
		\item \emph{Traceability}:
		The authorized entity must be able to obtain the real identity from a pseudo-identity by tracking the identity of an abnormal entity.
		\item \emph{Fine-grained Access Control}:
		If fine-grained access control is not achieved, unauthorized entities may be able to access sensitive data, leading to security breaches, data leaks, and other security threats. Therefore, it is necessary to implement fine-grained access control mechanisms to prevent those security threats.
		\item \emph{Policy Privacy}:
		To prevent potential analysis or attacks that could interfere with normal communication, the policy of the message must not be exposed during the encryption and decryption procedures.
	\end{itemize}
	
	\subsection{Security Definitions of OEEP-ABE and OE-IBS}\label{subsec:Securitydef1}
	This section defines the security properties to prove the security of the outsourcing-enabled and enhanced privacy-preserving ABE (OEEP-ABE) and outsourcing-enabled IBS (OE-IBS), proposed in Sec.~\ref{sec:Preliminary}. The \ac{OEEP-ABE} and \ac{OE-IBS} are the variants of PEAPOD \cite{KapTsaSmi:07} and online/offline IBS \cite{LjkBjZj:10}, respectively, with the properties of computation outsourcing and hidden policy. Thus, the following security definitions are based on \cite{KapTsaSmi:07} and \cite{LjkBjZj:10}.
	
	\begin{definition}[Ciphertext indistinguishability against chosen plaintext attack
		and restricted user collusion attack (C-IND-CPA-RUCA)] \label{def:1:C-IND-CPA-RUCA}
		A \ac{PPT} adversary $\mathcal{A}$ against the confidentiality of the OEEP-ABE can only corrupt either of one SA or any users (\textbf{Restriction 1:} Any collusions of corrupted users do not collectively satisfy\footnote{Here, ``collectively satisfy policy $P$" refers to a case where a combination of attributes, possessed by colluding the users, satisfies the policy \cite{KapTsaSmi:07}.} the target policy $P^*$, and SAs do not collude with each other.). $\mathcal{A}$ can learn their secrets, and issues requests for the output of \textbf{Out.Encrypt1} (i.e., $MO_C$ or $EM_C$) in polynomial times. $\mathcal{A}$ produces two equal-length messages $M_0$ and $M_1$ and a policy $P^*$. A challenger $\mathcal{C}$ flips a random coin $b \stackrel{R}{\in} \{0,1\}$ and generates $C^*$ by encrypting the message $M_b$ under the policy $P^*$ by \textbf{Encrypt} or \textbf{Select.Policy} that uses the issued $MO_C$ or $EM_C$ according to $b$. $\mathcal{C}$ returns $C^*$ to $\mathcal{A}$.
		
		The OEEP-ABE is said to be C-IND-CPA-RUCA if $\mathcal{A}$ can distinguish a ciphertext $C^*$, which is the encryption for one of $M_0$ and $M_1$, with only a negligible advantage.
	\end{definition}
	
	\begin{definition}[Policy indistinguishability against chosen plaintext attack
		and restricted user collusion attack (P-IND-CPA-RUCA)]\label{def:2:P-IND-CPA-UCA}
		$\mathcal{A}$ against the policy privacy of the OEEP-ABE can only corrupt either of one SA or any users (\textbf{Restriction 2:} All corrupted users do not satisfy any of the policies, $P_0$ and $P_1$ or satisfy both of the policies, and SAs do not collude with each other). $\mathcal{A}$ can learn their secrets, and issues requests for the outputs of \textbf{Out.Encrypt1} (i.e., $MO_C$ or $EM_C$) in polynomial times. $\mathcal{A}$ produces a chosen message $M^*$, and the two chosen policies, $P_0$ and $P_1$. A challenger $\mathcal{C}$ flips a random coin $b \stackrel{R}{\in} \{0,1\}$ and encrypts $M^*$ by \textbf{Encrypt} or \textbf{Select.Policy} that uses the issued $MO_C$ or $EM_C$ according to $b$. $\mathcal{C}$ returns $C^*$ to $\mathcal{A}$.
		
		The OEEP-ABE is said to be P-IND-CPA-RUCA if $\mathcal{A}$ can distinguish a ciphertext $C^*$, which is encryption of $M^*$ with one of $P_0$ and $P_1$, with only a negligible advantage.
	\end{definition}

	\begin{definition}[Unforgeability of IBS under chosen message attack (UF-IBS-CMA)]\label{def:3:UF-IBS-CMA}
	$\mathcal{A}$ against the unforgeability of the OE-IBS can issue a private key extraction query for an identity $ID$, and an adversary $\mathcal{A}$ can issue \textbf{Out.Sign1} queries with randomly selected $x$ for outsourced signing. A challenger then runs \textbf{KeyGen} providing $ID$ as an input, and obtains a corresponding private key $SSK_{ID}$. It also runs \textbf{Out.Sign1} using $MSPK$ and $x$, obtains signature materials $Y^x$, and gives them to $\mathcal{A}$. Note that, $Y$ is $g^y$, where $y$ is a randomly selected value. $\mathcal{A}$ can issue a signature generation query for $ID$, a message $M$, and $Y^x$. The challenger then runs \textbf{Sign} or \textbf{Out.Sign2} providing $SSK_{ID}$, $M$, and $Y^x$ as inputs, and then gives a generated signature $\sigma$ to $\mathcal{A}$. $\mathcal{A}$ outputs $ID'$, $M'$, $\sigma'$ (\textbf{Restriction 3:} $ID'$ and $M'$ have not been used in previous private key extraction and signature generation queries).
	$\mathcal{A}$ wins the security game if $\sigma '$ is a valid signature of $M'$. $\mathcal{A}$ is said to be an ($\epsilon$, $t$, $q_e$, $q_s$, $q_h$) forger if $\mathcal{A}$ has advantage at least $\epsilon$ in the above game, runs in time at most $t$, and makes at most $q_e$, $q_s$ and $q_h$. Note that, $q_e$, $q_s$ and $q_h$ are private key extraction, signing, and random oracle queries, respectively. 	
	
	The OE-IBS is said to be ($\epsilon$, $t$, $q_e$, $q_s$, $q_h$)-secure in the sense of UF-IBS-CMA if no ($\epsilon$, $t$, $q_e$, $q_s$, $q_h$)-forger exists.
	\end{definition}
\vspace{-2mm}
	\subsection{Security Definitions of PS-E2EID}\label{subsec:Securitydef2}
	This section captures the capabilities of attackers based on the definitions in our system and attack models. We first describe our notations in the security model. Our proposed protocol is denoted as $\Pi$, and communications between two users $A$ and $B$ in communication sessions of $t_1$ and $t_2$ are denoted as $\Pi_{A,B}^{t_1}$ and $\Pi_{B,A}^{t_2}$, respectively. Furthermore, we explain oracles and consider attackers who can query the oracles. We then define the security of our proposed protocols according to security requirements. The oracles, used to consider the attacker's capabilities, are as follows.
	\begin{itemize}
		\item \textbf{Execute}($\Pi_{A,B}^{t_1}$, $\Pi_{B,A}^{t_2}$) : This oracle models all types of passive attacker eavesdropping all data exchanged between $\Pi_{A,B}^{t_1}$ and $\Pi_{B,A}^{t_2}$.
		
		\item \textbf{Send}($\Pi_{A,B}^{t_1}$, $M$) : This oracle models an active attacker transmitting a message $M$ to $\Pi_{A,B}^{t_1}$.
		
		\item \textbf{Expose}($\Pi_{A,B}^{t_1}$) : This oracle models the exposure of the session key of $A$ shared with $B$ in a session $t_1$.
		
		\item \textbf{Corrupt}($\Pi_{A,B}^{t_1}$) : This oracle models the exposure of the long-term secret key of $A$ shared with $B$ in a session $t_1$.
		
		\item \textbf{CorruptAK}($S_i$) : This oracle models the exposure of an attribute-based private key $SK_i$ corresponding to an attribute set $S_i$. 
		
		\item \textbf{Test}($\Pi_{A,B}^{t_1}$): This oracle models a test of session key security. When one queries this oracle, it returns a real session key or a random string according to a random bit $b\in\{0,1\}$, if both entities, which are the partners of each other in the protocol, are accepted. Otherwise, it returns an invalid output.
		
		\item \textbf{TestPolicy}($\Pi_{A,B}^{t_1}, P_0, P_1, M$) This oracle models a test of policy privacy. When one queries this oracle with the inputs of two given policies, respectively denoted as $P_0$ and $P_1$, and a message $M$, it produces an encryption for $M$ with $P_b$, depending on randomly selected $b\in\{0,1\}$.
		
		\item \textbf{TestID}($\Pi_{A,B}^{t_1}$) This oracle models the test of identity anonymity. When one queries this oracle, it returns a real identity of $A$ or a random string, depending on random bit $b\in\{0,1\}$, if both parties, which are the partners of each other in the protocol, are accepted. Otherwise, it returns an invalid output.
		
	\end{itemize}
	
We define the security properties of the practical and secure vehicular communication protocol for E2E security to in-vehicle end-devices (PS-E2EID), according to the security requirements in Sec. II-A. We give the security definitions based on \cite{MbPr:93,McCbJm:10}, some of which are newly given based on the security requirements in our paper.

	\begin{definition}[Message Authentication] \label{def:5:Message-authentication}
	A protocol $\Pi$ is said to be with secure message authentication if any \ac{PPT} adversary $\mathcal{A}$ can only obtain a negligible probability to win the following security game.
	\begin{itemize}
		\item \textbf{Setup Phase} : A pair of functions, $\text{MF}$ and $\text{VF}$, are generated as the signature generation and verification functions of the message authentication, respectively, where $\sigma=\text{MF}(M)$ and (true,false) = $\text{VF}(\sigma,M)$. If $\sigma$ is generated by a specific $\text{MF}$ for $M$, the output of the corresponding $\text{VF}$ is true for the inputs of the corresponding $\sigma$ and $M$. Otherwise, it is false.
		\item \textbf{Initialization Phase} : $\text{VF}$ and all of the related public parameters are given to $\mathcal{A}$.
		\item \textbf{Training Phase} : $\mathcal{A}$ is allowed to request for the corresponding $\sigma_{q_1}, ..., \sigma_{q_n}$ for the selected $M_{q_1}, ..., M_{q_n}$.
		\item \textbf{Challenge Phase} : $\mathcal{A}$ finally outputs $\sigma^*$ for the corresponding selected $M^*$. $\mathcal{A}$ is said to win this game, if $\text{VF}(\sigma^*, M^*)$ = true, where $M^* \notin {M_{q_1}, ..., M_{q_n}}$.
	\end{itemize}
	If $\mathcal{A}$ wins the game, it has the following advantage:  
	\begin{align}
		\text{Adv}_{\mathcal{A}}^{\text{MeAuth}}=\text{Pr}[\mathcal{A}\ \text{win this security game}].
	\end{align}
\end{definition}

	\begin{definition}[Attribute-based Key Exchange] \label{def:6:Attribute-based-Key-exchange}
		$\mathcal{A}$ can query \textbf{Execute}, \textbf{Send}, and \textbf{CorruptAK} oracles in polynomial time. After this phase, $\mathcal{A}$ can initiate an instance $\Pi_{A,B}^{t_1}$ with a selected policy that cannot be satisfied by the attributes of the secret keys from \textbf{CorruptAK}. $\mathcal{A}$ then queries \textbf{Test} to gain a session key $K$ or a random string, according to random bit $b \in \{0,1\}$. After that $\mathcal{A}$ produces a guess $b'\in\{0,1\}$. If $b=b'$, $\mathcal{A}$ wins. Here, the advantage of $\mathcal{A}$ to break this security game is defined as
				\vspace{-1mm}
		\begin{align}
		\text{Adv}_{\mathcal{A}}^{\text{AKE}}=\text{Pr}[\text{Succ}_{\mathcal{A}}^{\text{AKE}}]\ -\ 1/2,
		\end{align}
		where $\text{Succ}_{\mathcal{A}}^{\text{AKE}}$ is the event that $\mathcal{A}$ produces a guess $b' = b$. If $\text{Adv}_{\mathcal{A}}^{\text{AKE}}$ is negligible, the attribute-based key exchange security is achieved.
	\end{definition}	

	\begin{definition}[Policy Privacy] \label{def:7:Policy-Privacy}
		A simulator $S$ simulates $\Pi_{A,B}^{t_1}$, and interacts with $\mathcal{A}$, who can query polynomial numbers of \textbf{Execute} and \textbf{Send} oracles in polynomial time. After this phase, $\mathcal{A}$ queries \textbf{TestPolicy} with a message $M$ and two valid policies, $P_0$ and $P_1$ as inputs to obtain $C_0$ or $C_1$ according to a random bit $b \in \{0,1\}$, where $C_0$ and $C_1$ are encryption for $M$ with $P_0$ and $P_1$, respectively. $\mathcal{A}$ has the following advantages:
				\vspace{-1mm}
		\begin{align}
			\text{Adv}_{\mathcal{A}}^{\text{PP}}=\text{Pr}[\text{Succ}_{\mathcal{A}}^{\text{PP}}]\ -\ 1/2,
		\end{align}
		where $\text{Succ}_{\mathcal{A}}^{\text{PP}}$ is the event that $\mathcal{A}$ produces a guess $b' = b$.
		If $\text{Adv}_{\mathcal{A}}^{\text{PP}}$ is negligible, the policy privacy is achieved.
	\end{definition}
	
	\begin{definition}[Identity Anonymity] \label{def:8:Identity-Anonymity}
		$S$ simulates $\Pi_{A,B}^{t_1}$, and interacts with $\mathcal{A}$, who can query polynomial numbers of \textbf{Execute} and \textbf{Send} oracles in polynomial time. After this phase, $\mathcal{A}$ queries \textbf{TestID} to obtain $\text{ID}_A$ or a random string from $\Pi_{A,B}^{t_1}$ according to a random bit $b \in \{0,1\}$. $\mathcal{A}$ has the following advantage:
				\vspace{-1mm}
		\begin{align}
		\text{Adv}_{\mathcal{A}}^{\text{ID-an}}=\text{Pr}[\text{Succ}_{\mathcal{A}}^{\text{ID-an}}]\ -\ 1/2,
		\end{align}
		where $\text{Succ}_{\mathcal{A}}^{\text{ID-an}}$ is the event that $\mathcal{A}$ produces a guess $b' = b$.
		If $\text{Adv}_{\mathcal{A}}^{\text{ID-an}}$ is negligible, the identity anonymity is achieved.
	\end{definition}


In our paper, we use the hash function (e.g., SHA-256) and the symmetric encryption (e.g., AES128 and AES256) as the pseudorandom function \cite{MbRcHk:96} and the pseudorandom permutation \cite{MlCr:88}, respectively. Therefore, we need their security definitions. However, we do not provide their definitions since their definitions are well-known.

	\vspace{-1mm}
	\section{Preliminaries}\label{sec:Preliminary}
	This section explains the intuitions and the constructions of the proposed OEEP-ABE and the OE-IBS.
	\vspace{-1mm}
	\subsection{Outsourcing-Enabled and Enhanced Privacy-preserving ABE (OEEP-ABE)}\label{subsec:OEEPABE}
	\subsubsection{Intuition}
	The \ac{OEEP-ABE} is a variant of PEAPOD \cite{KapTsaSmi:07} and EABEHP \cite{YuHsuLeeLee:20}. Compared to traditional ABEs, the \ac{OEEP-ABE} additionally supports outsourced encryption operations and a lightweight decryption algorithm, compared to traditional ABEs. Besides, \ac{OEEP-ABE} achieves confidentiality and policy privacy against outsider attackers and honest-but-curious \acp{SA} which perform outsourced encryption.
	
	The entire encryption procedure of OEEP-ABE consists of three functions: \textbf{Out.Encrypt1}, \textbf{Out.Encrypt2}, and \textbf{Select.Policy}. To avoid the exposure of the secrets of ciphertext decryption, the two distinct non-colluding SAs respectively generate different partial ciphertexts. Here, each of the two partial ciphertexts (i.e., $MO_{\tau}$ and $MO_{\tau '}$) is generated by the $\tau$-th and $\tau '$-th queries for \textbf{Out.Encrypt1}, respectively. The two partial ciphertexts $MO_{\tau}$ and $MO_{\tau '}$, from the non-colluding SAs are collected by the sender. A preliminary ciphertext $pC$ is then generated by the sender using the partial ciphertexts through \textbf{Out.Encrypt2}. Finally, a complete ciphertext $C$ is generated using the preliminary ciphertext $pC$ through \textbf{Select.Policy}. Here, \textbf{Out.Encrypt2} and \textbf{Select.Policy} only require multiplicative operations, considered as lightweight.
	
	The decryption function of OEEP-ABE only takes the number of multiplicative operations, proportional to the number of attributes and two exponential operations. In most of the previous ABE schemes, they require pairing operations and/or exponential operations whose number increases linearly with the number of attributes. On the other hand, the computation cost of the decryption function in the OEEP-ABE is considerably lower than those in the other ABE schemes.	
	\subsubsection{Construction}
	The \ac{OEEP-ABE} consists of 7 algorithms, which contain lightweight decryption and outsourced encryption. The construction of the algorithms is shown as follows.
	\begin{itemize}
		\item \textbf{Setup($1^{\lambda}$)}: This algorithm selects a cyclic group $\mathbb{G}$ of prime order $p$ with a generator $g$. A large prime number $q$ is then selected such that $q|(p-1)$. It generates $\{a_i\}_{i\in{}I} \in \mathbb{Z}^{*}_q$ for all $i \in I$ and $d \in \mathbb{Z}^{*}_q$, where $I=\{1, 2, 3, \ldots , N\}$ is the universal set of attribute indices of the system, and $N$ is the number of system attributes. The master public key $MPK$ and the master secret key $MSK$ of the system are then generated as
				\vspace{-1mm}
		\begin{align} \notag
			&MPK=\{g, \; p, \; q, \; g^d , \{ PK_{S_i}=g^{a_i}\}_{i\in{}I}\}, \notag \\
			&MSK = \{d, \{a_i\}_{i\in I} \}. \notag
		\end{align}
		
		\item \textbf{KeyGen}($MSK,\text{ID}_j,I_j$): This algorithm randomly selects $\{a_{i,j,1} , a_{i,j,2}\}_{i \in I_j} \in (\mathbb{Z}^{*}_q)^2$ for all $i \in I_j$ and $s_j \in \mathbb{Z}^{*}_q$ such that $\sum_{i \in I_j} (a_{i,j,1} + a_{i,j,2}) = \sum_{i \in I_j} a_i$. Here, $I_j \subseteq I$ is the set of attributes indices of user $j$. It then generates user secret key as 
				\vspace{-1mm}
		\begin{align}
			SK_{U_j} &=\{SK_{U_{j,1}} , SK_{U_{j,2}}\} \notag \\
			&= \left\{s_j + \sum_{i \in I_j} a_{i,j,1}  \; , \; (- s_j + \sum_{i \in I_j} a_{i,j,2}) \cdot d^{-1}\right\} \notag
		\end{align}
			
		\item \textbf{Encrypt($MPK, T, M$)}: This algorithm first takes $MPK$, $T$, and $M$ as inputs, and produces the ciphertext, $C$, on $M$. The policy set $T=\{t_i\}_{i\in I}$ is determined as $t_i = 1$ if the attribute $i$ is required, $t_i = 0$ if the attribute $i$ is irrelevant, and $t_i = -1$ if the attribute $i$ is forbidden. When $M\in \mathbb{Z}_q$ is the message to be encrypted, message tuples are then generated for each $i\in{}I$ as
				\vspace{-1mm}
		\begin{equation}
			p_i=
			\begin{cases} k_i & \text{if} \; t_{i}=1
				\\ 1 & \text{if} \; t_{i}=0
				\\ \alpha_i & \text{if} \; t_{i}=-1
			\end{cases}\nonumber 	\;		\text{s.t.} \prod_{\substack {t_i\in T \wedge t_i = 1, \\ \forall i\in I}} \hspace{-0.4mm} p_i\equiv M \; (mod \; q),
		\end{equation}
		for randomly selected $\alpha_i\in \mathbb{Z}^{*}_q$. Next, it randomly selects $r\in \mathbb{Z}^{*}_q$ and encrypts each message tuple, denoted by $p_i$, using $PK_{S_i} = g^{a_i}$ as
				\vspace{-1mm}
		\begin{equation}
			C= \{A, \{B_i\}_{i \in I} , D \}=\{g^{r}, \{p_i (g^{a_i})^{r}\}_{i \in I} , (g^d)^{r} \}\nonumber
		\end{equation}

		\item \textbf{Out.Encrypt1($MPK, v_{\tau}$)}: This algorithm takes $MPK$ and $v_{\tau}$ as inputs and produces $MO$ as
		\vspace{-1mm}
		\begin{align} 
			&MO = \{MO_{\tau, 1} , MO_{\tau, 2} , MO_{\tau, 3}\} \notag \\
			& = \{g^{v_{\tau}} , \{PK^{v_{\tau}}_{S_i}\}_{i \in I} , (g^d)^{v_{\tau}}\} = \{g^{v_{\tau}} , \{g^{a_i v_{\tau}}\}_{i \in I} , g^{dv_{\tau}}\}, \notag
		\end{align}
		where $v_{\tau}$ is randomly selected from $\mathbb{Z}^{*}_q$ at the $\tau$-th query.
		
		\item \textbf{Out.Encrypt2($MPK, MO_{\tau}, MO_{\tau '}$)}: This algorithm takes $MPK$, $MO_{\tau}$, and $MO_{\tau '}$ as inputs, and produces the preliminary ciphertext, $pC$. Here, $\tau$ and $\tau '$ are the indices of the queries for producing $MO_{\tau}$ and $MO_{\tau '}$ in \textbf{Out.Encrypt1}. We denote $r = v_{\tau}+v_{\tau '}$. It then computes $pC$ as
		\vspace{-1mm}
		\begin{align} \notag
			pC &\hspace{-0.4mm}=\hspace{-0.4mm} \{\hspace{-0.5mm}MO_{\tau, 1} \hspace{-0.6mm}\cdot \hspace{-0.6mm} MO_{\tau '\hspace{-0.5mm}, 1},\hspace{-0.2mm} MO_{\tau, 2} \hspace{-0.6mm}\cdot \hspace{-0.6mm} MO_{\tau '\hspace{-0.5mm}, 2},\hspace{-0.2mm} MO_{\tau, 3} \hspace{-0.6mm}\cdot \hspace{-0.6mm} MO_{\tau '\hspace{-0.5mm}, 3}\hspace{-0.4mm}\} \notag \\ 
			&\hspace{-0.4mm}=\hspace{-0.4mm} \{g^{v_{\tau}} \hspace{-0.4mm}\cdot \hspace{-0.4mm} g^{v_{\tau '}} , \{PK^{v_{{\tau}}}_{S_i}\hspace{-0.4mm} \cdot \hspace{-0.4mm}PK^{v_{{\tau '}}}_{S_i}\}_{i \in I} , (g^d)^{v_{{\tau}}}\hspace{-0.4mm} \cdot \hspace{-0.4mm}(g^d)^{v_{{\tau '}}}\} \notag \\
			&\hspace{-0.4mm}=\hspace{-0.4mm} \{g^{r}, \{g^{a_i r}\}_{i \in I} , g^{dr}\} = \{A, \{B'_i\}_{i \in I} , D\} \notag
		\end{align}
	
		\item \textbf{Select.Policy($pC, T, M$)}: This algorithm takes $pC$, $T$, and $M$ as inputs, and produces the ciphertext, $C$, on $M$. Message tuples are then generated using the same method described in \textbf{Encrypt}. It encrypts each of the message tuples, denoted as $p_i$, as
				\vspace{-1mm}
		\begin{align}
			C = \{A, \{p_i \cdot B'_i \}_{i \in I}, D\} = \{g^{r}, \{p_i g^{a_i r}\}_{i \in I} , \; g^{dr}\} \notag
		\end{align}	
		
		\item \textbf{Decrypt($C, SK_{U_r} , I_r$)}: This algorithm takes $C$, $SK_{U_r}$, and $I_r$ as inputs, and produces $M$ as
		\vspace{-1mm}
		\begin{equation}
			M = \prod_{i \in I_r} B_i / (A^{SK_{U_{r,1}}} \cdot D^{SK_{U_{r,2}}}) \notag
		\end{equation}
	\end{itemize}
	\vspace{-5mm}
	\subsection{Outsourcing-Enabled IBS (OE-IBS)}\label{subsec:OEIBS}
	\subsubsection{Intuition}
	The OE-IBS is identical to the signature scheme in \cite{LjkBjZj:10}, except for the outsourced signing. The OE-IBS has additional algorithms for outsource signing to the honest-but-curious SA. The outsourced signing consists of \textbf{Offline.Sign}, \textbf{Out.Sign2}, which are performed by a user, and \textbf{Out.Sign1}, which is performed by the SA. Similarly to the OEEP-ABE, the OE-IBS can selectively use outsourcing techniques since it can provide both the normal signing and the outsourced signing.
	\subsubsection{Construction}
	The OE-IBS consists of 7 algorithms including the normal signing and outsourced signing algorithms.
	\begin{itemize}
		\item \textbf{Setup($1^{\lambda}$)}: This algorithm selects a cyclic group $\mathbb{G}$ of the prime order $p$ with a generator $g$. It selects $x \in \mathbb{Z}^{*}_p$ and $\text{H}_1 : \{0,1\} \rightarrow \mathbb{Z}^{*}_p$. Then, it computes $X=g^x$. The master signature public key $MSPK$ and the master signature secret key $MSSK$ of the system are then generated as
				\vspace{-1mm}
		\begin{align} \notag
			MSPK=\{X , \text{H}_1 \} , \; MSSK = \{x\}
		\end{align}
		
		\item \textbf{KeyGen}($MSPK, MSSK,\text{ID}_j$): This algorithm randomly selects $\beta_j \in \mathbb{Z}^{*}_p$ and computes $\mathcal{B}_j=g^{\beta_j}$ and $\kappa_j = \beta_j + \text{H}_1(\mathcal{B}_j, \text{ID}_j)x$ mod $p$. The signature secret key of user $j$ $SSK_j$, is generated as $SSK_j = (\mathcal{B}_j, \kappa_j)$.
		
		\item \textbf{Sign($MSPK, SSK_j, M_t$)}: This algorithm randomly selects $y_t \in \mathbb{Z}^{*}_p$ and computes $Y_t = g^{y_t}$, $h_t=\text{H}_1(Y_t, \mathcal{B}_j, M_t)$, and $z_t = y_t + h_t \kappa_j$ mod $p$.
		The signature $\text{Sig}_{j,t}$ is generated as $\text{Sig}_{j,t} = (Y_t, \mathcal{B}_j, z_t)$.
		
		\item \textbf{Offline.Sign($MSPK$)}: This algorithm randomly selects $y, \omega \in \mathbb{Z}^{*}_p$ and computes $Y = g^{y}, \; g^{\frac{1}{\omega}}$.
		
		\item \textbf{Out.Sign1($Y$)}: This algorithm randomly selects $x_t \in \mathbb{Z}^{*}_p$ and computes $Y'_t= Y^{x_t}$. 
		
		\item \textbf{Out.Sign2($MSPK, SSK_j, g^{\frac{1}{\omega}}, Y'_t, M_t$)}: This algorithm computes $h_t =\text{H}_1(Y'_t, \mathcal{B}_j, M_t)$, and $z_t = x_t y + h_t\kappa_j$ mod $p$. After that it randomly selects $\mathcal{X}_t$ and $\mathcal{Y}_t$ that satisfy $z_t = \mathcal{X}_t + \mathcal{Y}_t$ mod $p$.
		The signature $\text{Sig}_{j,t}$ is generated as $\text{Sig}_{j,t} = (Y'_t, \mathcal{B}_j, \omega\mathcal{X}_t, g^{\frac{1}{\omega}}, \mathcal{Y}_t)$.
		
		\item \textbf{Verify}($MSPK, \text{ID}_j, \text{Sig}_{j,t}, M_t$): This algorithm takes $MSPK$ and $\text{Sig}_{j,t}$ as inputs and outputs 1 if the signature is valid or 0 otherwise. It first computes $h_t=\text{H}_1(Y_t, \mathcal{B}_j, M_t)$ or $h_t =\text{H}_1(Y'_t, \mathcal{B}_j, M_t)$ and $g^{z_t} = (g^{\frac{1}{\omega}})^{\omega\mathcal{X}_t} \cdot g^{\mathcal{Y}_t}$. Then, it verifies
				\vspace{-1mm}
		\begin{align} \notag
			g^{z_t} \stackrel{?}{=}
			\begin{cases} Y_t \mathcal{B}^{h_t}_j X^{h_t \text{H}_1(\mathcal{B}_j, \text{ID}_j)} & \text{if} \; \text{no outsourcing}
			\\ Y'_t \mathcal{B}^{h_t}_j X^{h_t \text{H}_1(\mathcal{B}_j, \text{ID}_j)} & \text{if} \; \text{outsourcing}.
			\end{cases}
		\end{align}
	If the equality holds, it outputs 1, otherwise it outputs 0.
	
	\item \textbf{BatchVerify}($MSPK, \{\text{ID}_j\}_{j\in J}, \{\text{Sig}_{j,t}, M_{j,t}\}_{j\in J, t\in \mathbb{T}}$): This algorithm takes $MSPK$ and $\{\text{Sig}_{j,t}\}_{j\in J, t\in \mathbb{T}}$ as inputs and outputs 1 if the signature is valid or 0 otherwise, where $\mathbb{T}$ is a set of indices of messages to be verified. It first computes $h_{j,t}=\text{H}_1(Y_{j,t}, {\mathcal{B}_j}, M_{j,t})$ or $h_{j,t} =\text{H}_1(Y'_{j,t}, \mathcal{B}_j, M_{j,t})$ for all $j\in J$ and $t\in \mathbb{T}$. In addition, it computes 
			\vspace{-1mm}
	\begin{align} \notag
	g^{\sum_{j\in J, t\in \mathbb{T}} z_{j,t}} = \prod_{j\in J}\bigg((g^{\frac{1}{\omega_j}})^{\sum_{t\in\mathbb{T}}\omega_j\mathcal{X}_{j,t}} \cdot g^{\sum_{t\in\mathbb{T}}\mathcal{Y}_{j,t}}\bigg).
	\end{align}
	Then, it verifies
			\vspace{-1mm}
	\begin{align} \notag
		g^{\sum_{j\in J, t\in \mathbb{T}} z_{j,t}} \stackrel{?}{=}
		\begin{cases} (\hspace{-2mm}\prod\limits_{j\in J, t\in \mathbb{T}}\hspace{-1.5mm} Y_{j,t} \mathcal{B}^{h_{j,t}}_j) X^{\sum_{j\in J, t\in \mathbb{T}}{h_{j,t}} \text{H}_1(\mathcal{B}_j, \text{ID}_j)} 
			\\ (\hspace{-2mm}\prod\limits_{j\in J, t\in \mathbb{T}}\hspace{-1.5mm} Y'_{j,t} \mathcal{B}^{h_{j,t}}_j) X^{\sum_{j\in J, t\in \mathbb{T}}{h_{j,t}} \text{H}_1(\mathcal{B}_j, \text{ID}_j)}
		\end{cases}
	\end{align}
	If the equality holds, it outputs 1, otherwise it outputs 0.
	\end{itemize}
	\vspace{-3mm}
	\section{Practical and Secure Vehicular Communication Protocol for End-to-End Security to In-vehicle End-devices (PS-E2EID)}\label{sec:protocol}
	The \ac{PS-E2EID} consists of four phases, which will be explained in the following subsections. Here, we denote as \emph{starting state} the state when the CAV is initially turned on, and as \emph{driving state} the state when the CAV is driven. In our system model, \acp{CAV}, \acp{RSU}, \acp{UE}, and \acp{OEM} periodically perform 5G-\ac{AKA} with the core network. The period of performing 5G-\ac{AKA} is different for each device, which is properly determined to maintain the security of our system. In addition, each entity receives an encryption key $CK$, an integrity key $IK$, an anonymity key $AK$ from UDM/ARPF via 5G-\ac{AKA}. In the following subsections, we explain each phase in detail. Note that, for simplicity, we consider the \ac{ADAS} of the $n$-th CAV $(\text{CAV}_n)$ as $\text{ECU}_{n0}$ in the proposed protocols.
	\vspace{-3mm}
	\subsection{System Initialization and Key Management}\label{subsec:InitKeymanage}
	This section presents how \ac{UDM}/\ac{ARPF} generate, distribute, and manage system parameters and keys for each entity. We then explain the method to generate and distribute encryption materials at OEM for each ECUs. \ac{UDM}/\ac{ARPF} generate $(MPK, MSK)$ and $(MSPK, MSSK)$ through \textbf{OEEP-ABE.Setup} and \textbf{OE-IBS.Setup}, respectively. They publish a master public key and master signature public key $(MPK, MSPK)$, and keep the master secret key and the master signature secret key $(MSK, MSSK)$. They generate and distribute a user secret key $SK_U$ and a user signature secret key $SSK_U$  by \textbf{OEEP-ABE.KeyGen} and \textbf{OE-IBS.KeyGen}, respectively, for each member of the system, except for \ac{OBU}. In addition, they generate and distribute long-term symmetric keys between \ac{ECU} and the \ac{SA} and between the \ac{ADAS} and the \ac{SA}, $K_{\text{SA},\text{ECU}_{nj}}$. Then, they generate and distribute long-term symmetric keys between the \ac{OEM} and \ac{ECU}, $K_{\text{OEM},\text{ECU}_{nj}}$. Finally, they generate their private key $x_{\text{CN}}$ and public key $y_{\text{CN}} = g^{x_{\text{CN}}}$, and publish the public key. 
	
	When an \ac{ECU} is manufactured, the \ac{OEM} securely installs the encryption materials $EM$ onto the \ac{ECU} using $K_{\text{OEM},\text{ECU}_{nj}}$. Here, the number of encryption materials can be properly decided by considering the storage space of the \ac{ECU} and the update period.
	The credentials, held by each device, are as follows.
	\begin{enumerate}
		\item The credentials of the $\text{(RSU/UE/OEM)}_m$ include a subscription concealed identifier $\text{SUCI}_m$ for the \ac{C-V2X} service, the public key $y_{\text{CN}}=g^{x_{\text{CN}}}$ of \ac{UDM}/\ac{ARPF}, $(MPK, MSPK)$, $SK_{U_m}$, and $SSK_{U_m}$. Note that $K_{\text{OEM},\text{ECU}}$ is additionally installed in the \ac{OEM}.
		
		\item  The credentials of the $\text{OBU}_n (\text{SA}_n)$ in $\text{CAV}_n$ include $\text{SUCI}_n$, $y_{\text{CN}}$, $(MPK, MSPK)$, $SSK_{U_n}$, and $\{K_{\text{SA},\text{ECU}_{nj}}\}_{j \in J_n}$, where $J_n = \{0,1, ..., j, ...\}$ is a set of indices of the \acp{ECU} and \ac{ADAS} mounted on $\text{CAV}_n$.
		
		\item The credentials of $\text{ECU}_{nj}$ of $\text{CAV}_n$ include $(MPK, MSPK)$, $SK_{U_{nj}}$, $SSK_{U_{nj}}$, $K_{\text{SA},\text{ECU}_{nj}}$, and $K_{\text{OEM},\text{ECU}_{nj}}$.
	\end{enumerate}
\vspace{-3mm}
	\begin{figure}[t!]
		\begin{center}
			
			{ 
				\includegraphics[width=0.99\columnwidth]{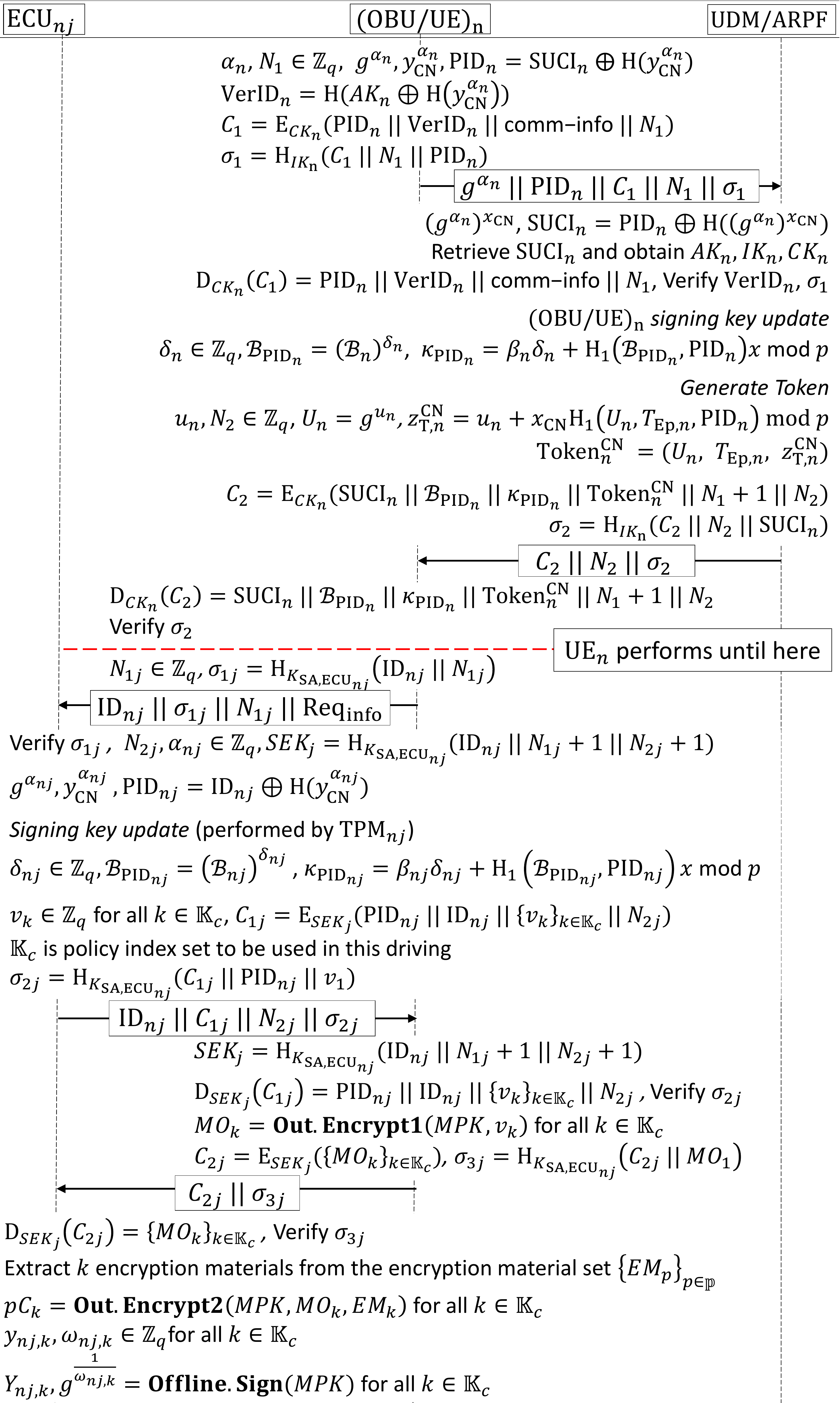}
			}
		\end{center}
		\caption{
			Initial authentication and preliminary protocol for OBU, UE, and ECUs.
			\vspace{-6mm}
		}
		\label{fig:preliminary}

	\end{figure}
	\subsection{Initial Authentication and Preliminary}
		Before the \ac{CAV} enters the \emph{driving state}, the \ac{CAV} performs a preliminary phase in the \emph{starting state}.\footnote{This phase can be performed not only between an \ac{ECU} and the \ac{SA}, but also between the \ac{SA} and a resource-poor external devices such as \ac{RSU} and \ac{UE}. In this case, the external device must have encryption materials, delivered from an authority that can securely provide them, such as UDM/ARPF. The external device can perform \textbf{Out.Encrypt1}, \textbf{Out.Encrypt2}, and \textbf{Select.Policy} online using those encryption materials.} This phase mainly includes the following three steps. 
		\begin{enumerate}
			\item Pseudo-identity generation: the OBU, UE, and ECU generate their own pseudo-identities and new signing keys, while the ones for the OBU and UE are generated by UDM/ARPF and the one for the ECU is by the \ac{TPM}, installed in the ECU. The OEM and RSU use their real identities without performing this step.
			\item Authentication token distribution: each entity performs authentication with UDM/ARPF, and then receives a token from UDM/ARPF to reduce the communications during data sharing. Thus, the \ac{CAV} can securely share data without communicating with UDM/ARPF unless the \ac{CAV} is restarted or the token is expired;\footnote{The periods of 1) and 2) can be changed according to the security level for unlinkability, required by the system. For example, 1) and 2) can be performed when the user moves to another area. Here, there can be a trade-off between the traffic with UDM/ARPF and the security level.} 
			\item Preliminary ciphertext and signature generation between the OBU and ECUs: the OBU and ECUs perform outsourced encryption, and the ECUs also perform offline signing to reduce the resources, consumed for encryption and signing in the attribute-based data sharing.
		\end{enumerate}
	
		This protocol, explained above, consists of the following 6 steps as shown Fig. \ref{fig:preliminary}. 

	\begin{enumerate}
		\item The $(\text{OBU/UE})_n$ randomly selects $\alpha_N$ and $N_1$, and generates $g^{\alpha_n}, y^{\alpha_n}_\text{CN}, \text{PID}_n = \text{SUCI}_n \oplus \text{H}(y^{\alpha_n}_\text{CN})$, and $\text{VerID}_n = \text{H}(AK_n \oplus \text{H}(y^{\alpha_n}_\text{CN})$. Then, it computes $C_1 = \text{E}_{CK_n}(\text{PID}_n || \text{VerID}_n || \text{comm-info} || N_1)$ and $\sigma_1 = \text{H}_{IK_n}(C_1 || N_1 || \text{PID}_n)$, and sends $g^{\alpha_n} || \text{PID}_n  || C_1 || N_1 \newline || \sigma_1$ to UDM/ARPF.
		
		\item The UDM/ARPF compute $(g^{\alpha_n})^{x_\text{CN}}$ and $\text{SUCI}_n$, retrieves $\text{SUPI}_n$, and obtain $AK_n, IK_n$, and $CK_n$. They then decrypt $C_1$ and verify $\text{VerID}_n, \sigma_1$. Then, it performs $(\text{OBU/UE})_n$ signing key update process by randomly selecting $\delta_n$ and generating new signing key $(\mathcal{B}_{\text{PID}_n} , \kappa_{\text{PID}_n})$. After that they randomly select $u_n$ and $N_2$, and generate an authentication token for $(\text{OBU/UE})_n$, $\text{Token}^{\text{CN}}_n = (U_n, T_{\text{Ep},n}, z^{\text{CN}}_{\text{T},n})$, as follows. They compute $U_n = g^{u_n}$ and $z^{\text{CN}}_{\text{T},n} = u_n + x_{\text{CN}}\text{H}_1(U_n, T_{\text{Ep},n}, \text{PID}_n) \;\text{mod}\; p$, where $T_{\text{Ep},n}$ is the token expiration time of $(\text{OBU/UE})_n$. Finally, they compute $C_2 = \text{E}_{CK_n}(\text{SUCI}_n || \mathcal{B}_{\text{PID}_n} || \kappa_{\text{PID}_n} || \text{Token}^{\text{CN}}_n || N_1 + 1 || N_2)$ and $\sigma_2 = \text{H}_{IK_n}(C_2 || N_2 || \text{SUCI}_n)$, and send $C_2 || N_2 || \sigma_2$ to $(\text{OBU/UE})_n$.
		
		\item The $(\text{OBU/UE})_n$ decrypts $C_2$ and verifies $\sigma_2$. Additionally, the $\text{OBU}_n$ randomly selects $N_{1j}$ and computes $\sigma_{1j} = \text{H}_{K_{\text{SA},\text{ECU}_{nj}}}(\text{ID}_{nj} || N_{1j})$. Then, it sends $\text{ID}_{nj} || \sigma_{1j} || N_{1j} || \text{Req}_\text{info}$ to $\text{ECU}_{nj}$.
		
		\item The $\text{ECU}_{nj}$ verifies $\sigma_{1j}$, randomly selects $N_{2j}, \alpha_{nj}$, and computes $SEK_j = \text{H}_{K_{\text{SA},\text{ECU}_{nj}}}(ID_j || N_{1j} + 1 || N_{2j} + 1)$. It also computes $g^{\alpha_{nj}}, y^{\alpha_{nj}}_\text{CN}$, and $\text{PID}_{nj} = \text{ID}_{nj} \oplus \text{H}(y^{\alpha_{nj}}_\text{CN})$. After that it performs the signing key update process through mounted $\text{TPM}_{nj}$, and obtains new signing key $(\mathcal{B}_{\text{PID}_{nj}} , \kappa_{\text{PID}_{nj}})$. Next, it randomly selects $v_{k}$ for all $k \in \mathbb{K}_c$, where $\mathbb{K}_c$ is the index set of message types (i.e., policies) to be used in this driving. Finally, it computes $C_{1j} = \text{E}_{SEK_j}(\text{PID}_{nj} || \text{ID}_{nj} || \{v_{k}\}_{k\in \mathbb{K}_c} || N_{2j})$ and $\sigma_{2j} = \text{H}_{K_{\text{SA},\text{ECU}_{nj}}}(C_{1j} || \text{PID}_{nj} || v_{1})$, and sends $\text{ID}_{nj} || C_{1j} || N_{2j} || \sigma_{2j}$ to $\text{OBU}_n$.
		
		\item The $\text{OBU}_n$ computes $SEK_j$ and decrypts $C_{1j}$. It then verifies $\sigma_{2j}$ and computes $MO_{k} = \text{\textbf{Out.Encrypt1}}\newline(MPK, v_{k})$ for all $k \in \mathbb{K}_c$. Next, it computes $C_{2j} = \newline \text{E}_{SEK_j}(\{MO_{k}\}_{k \in \mathbb{K}_c})$ and $\sigma_{3j} = \text{H}_{K_{\text{SA},\text{ECU}_{nj}}}(C_{2j} || \newline  MO_{1})$. Finally, it sends $C_{2j} || \sigma_{3j}$ to $\text{ECU}_{nj}$.
		
		\item The $\text{ECU}_{nj}$ decrypts $C_{2j}$ and verifies $\sigma_{3j}$. It then extracts $k$ encryption materials from the encryption material set $\{EM_{p}\}_{p \in \mathbb{P}}$, where $\mathbb{P}$ is the index set of unused encryption materals, held by the $\text{ECU}_{nj}$. Next, it computes $pC_{k} = \text{\textbf{Out.Encrypt2}}(MPK, MO_{k}, EM_{k})$, randomly selects $y_{nj,k}$ and $\omega_{nj,k}$, and computes $Y_{nj,k}, g^{\frac{1}{\omega_{nj,k}}} =  \text{\textbf{Offline.Sign}}(MPK)$ for all $k \in \mathbb{K}_c$.
	\end{enumerate}
\vspace{-3mm}
\begin{figure}[t!]
	\vspace{-6.5mm}
	\begin{center}
		
		{ 
			\includegraphics[width=1\columnwidth]{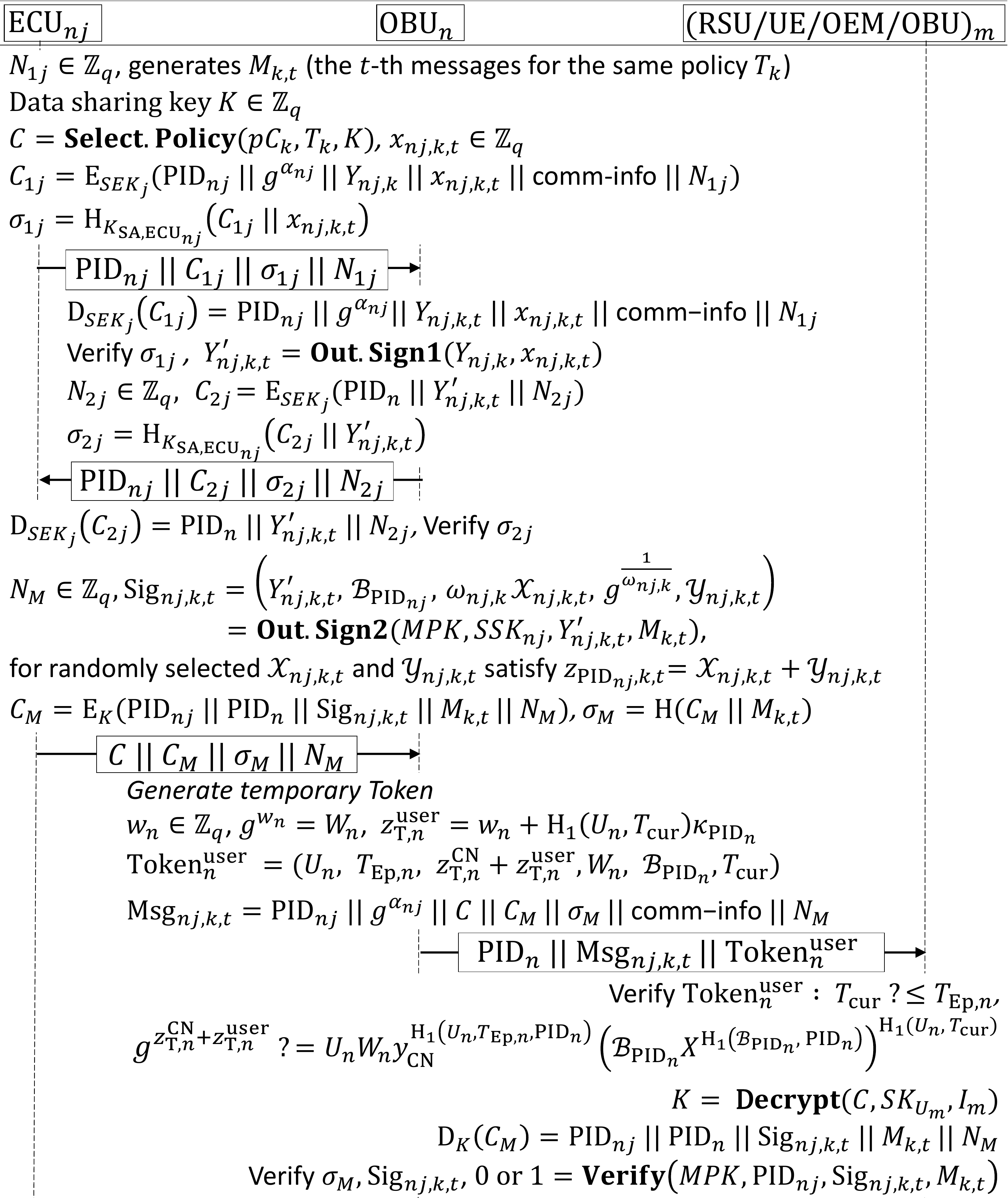}
		}
	\end{center}
	\caption{
		Uplink attribute-based data sharing (ABDS) protocol between the in-vehicle end-devices and the RSU.
		\vspace{-5mm}	
	}
	\label{fig:upABDS}
	
\end{figure}
	\subsection{Attribute-based Data Sharing (ABDS)}
	The ABDS protocol consists of the \textit{uplink} ABDS protocol to transfer data from in-vehicle end-devices to \ac{RSU}/\ac{UE}/\ac{OEM}/\ac{OBU} and the \textit{downlink} ABDS protocol to transfer data from \ac{RSU}/\ac{UE}/\ac{OEM} to in-vehicle end-devices.
	
\begin{figure}[t!]
			\vspace{-6.5mm}
		\begin{center}
			
			{ 
				\includegraphics[width=1\columnwidth]{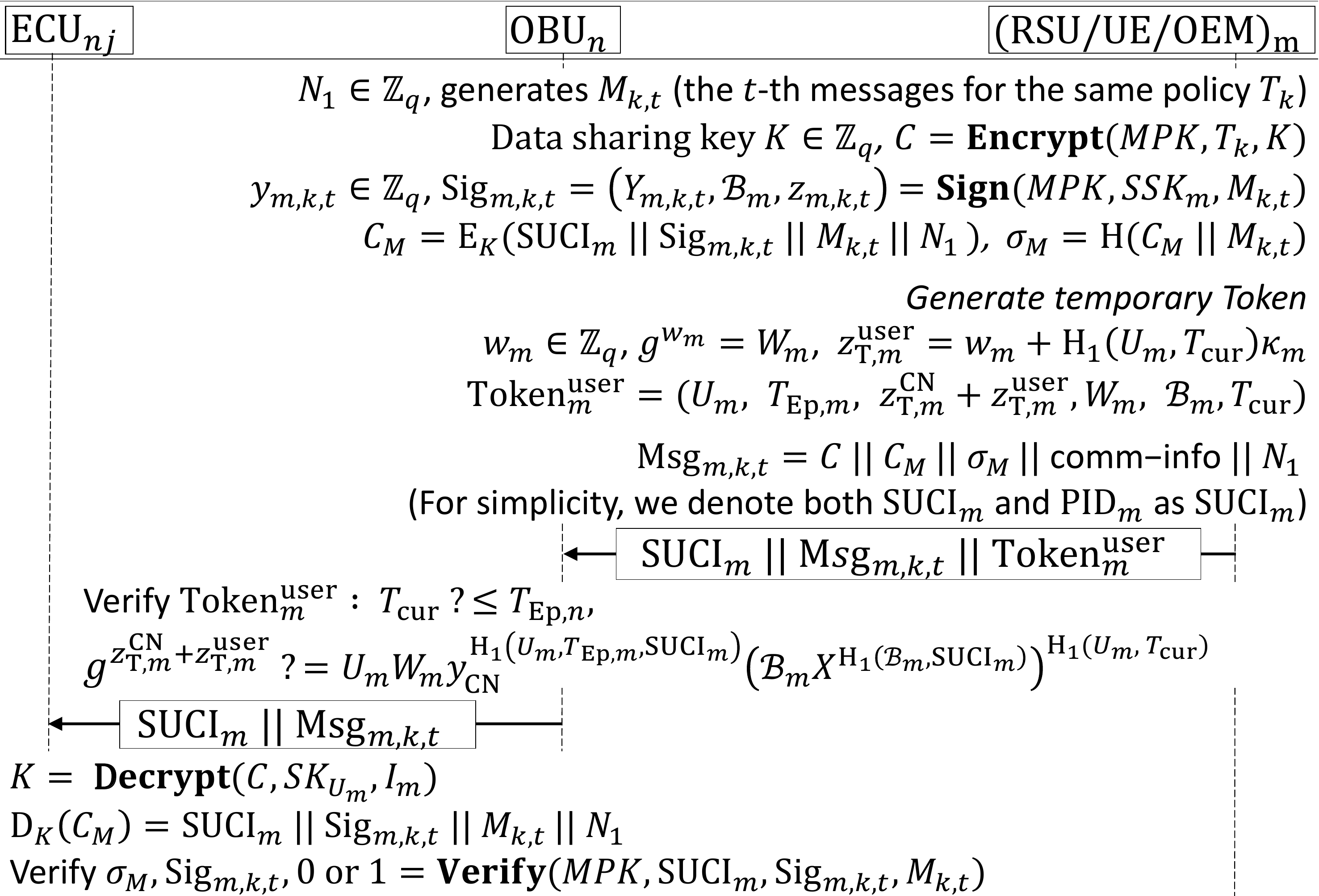}
			}
		\end{center}
		\caption{
			Downlink attribute-based data sharing (ABDS) protocol between a RSU and in-vehicle end-devices.
		}
		\label{fig:downABDS}
		\vspace{-6mm}
	\end{figure}

	In the uplink ABDS protocol, the \ac{ECU} can quickly compute a ciphertext using the preliminary ciphertext from the previous phase. In this protocol, we assume that the outsourced encryption and signing are used by the \ac{ECU}, and the normal encryption and signing are used by the other entities. In addition, a message receiver can validate the legitimacy of a sender based on the token (i.e., the OBU in case of the \ac{CAV}). The uplink and downlink ABDS protocols consist of 5 and 3 steps as shown in Figs.~\ref{fig:upABDS} and \ref{fig:downABDS}, respectively. To emphasize the objective of our research, only \acp{ECU} are included in the ABDS protocols without the ADAS. Note that we only include the description on the uplink ABDS protocol.
	
	\begin{enumerate}
		
		\item The $\text{ECU}_{nj}$ generates $N_{1j}$ and message $M_{k,t}$, where $M_{k,t}$ is $t$-th messages for the same policy $T_k$. It then generates data sharing key $K$ and computes $C = \text{\textbf{Select.Policy}}(pC_{k}, T_k, K)$. After that it randomly selects $x_{nj,k,t}$ and computes $C_{1j} = \text{E}_{SEK_j}(\text{PID}_{nj} || g^{\alpha_{nj}} || Y_{nj,k} || x_{nj,k,t} || \text{comm-info} || N_{1j})$ and $\sigma_{1j} = \text{H}_{K_{\text{SA},\text{ECU}_{nj}}}(C_{1j} || x_{nj,k,t})$. Finally, it sends $C_{1j} || \sigma_{1j} || N_{1j}$ to $\text{OBU}_n$.
		
		\item The $\text{OBU}_n$ decrypts $C_{1j}$ and verifies $\sigma_{1j}$. Next, it computes $Y'_{nj,k,t} = \text{\textbf{Out.Sign1}}(Y_{nj,k}, x_{nj,k,t})$ and randomly selects $N_{2j}$. Then, it computes $C_{2j} = \text{E}_{SEK_j}(\text{PID}_n || \newline Y'_{nj,k,t} || N_{2j})$ and $\sigma_{2j} = \text{H}_{K_{\text{SA},\text{ECU}_{nj}}}(C_{2j} || Y'_{nj,k,t})$, and sends $C_{2j} || \sigma_{2j} || N_{2j}$ to $\text{ECU}_{nj}$.
		
		\item The $\text{ECU}_{nj}$ decrypts $C_{2j}$ and verifies $\sigma_{2j}$. It then randomly selects $N_M$ and computes $\text{Sig}_{nj,k,t} = \text{\textbf{Out.Sign2}}(MPK, SSK_{nj}, Y'_{nj,k,t}, M_{k,t})$. Next, it computes $C_M = \text{E}_K(\text{PID}_{nj} || \text{PID}_n || \text{Sig}_{nj,k,t} || M_{k,t} || N_M)$ and $\sigma_M = \text{H}(C_M || M_{k,t})$. Finally, it sends $C || C_M || \sigma_M \newline|| N_M$ to $\text{OBU}_n$.
		
		\item The $\text{OBU}_n$ randomly selects $w_n$ and generate temporary authentication token $\text{Token}^{\text{user}}_n = (U_n, T_{\text{Ep},n}, z^{\text{CN}}_{\text{T},n}+z^{\text{user}}_{\text{T},n}, W_n, \mathcal{B}_{\text{PID}_n}, T_{cur})$, where $W_n = g^{w_n}, z^{\text{user}}_{\text{T},n} = w_n + \text{H}_1(U_n, T_{cur})\kappa_{\text{PID}_{nj}}$, and $T_{cur}$ is a timestamp for the current time. Finally, it sends $\text{PID}_n || \text{Msg}_{nj,k,t} || \text{Token}^{\text{user}}_n$ to $(\text{RSU/UE/OEM/OBU})_m$.
		
		\item The $(\text{RSU/UE/OEM/OBU})_m$ verifies $\text{Token}^{\text{user}}_n$ and obtains $K$ through $\text{\textbf{Decrypt}}(C, SK_{U_m}, I_m)$. Next, it decrypts $C_M$, and verifies $\sigma_M$, and $\text{Sig}_{nj,k,t}$ through $\text{\textbf{Verify}}(MPK,  \text{PID}_{nj}, \text{Sig}_{nj,k,t}, M_{k,t})$.
	\end{enumerate}
\vspace{-3mm}
	\subsection{Encryption Material Update}
	When an ECU runs out of the encryption materials, initially installed by the OEM, new encryption materials need to be re-installed onto the ECU. The update period is sufficiently long as the encryption material size is generally large, and it can be done at a vehicle service station.
\vspace{-3mm}
	\section{Security Analysis}\label{sec:securityanalysis}
	This section proves the security of the \ac{OEEP-ABE}, \ac{OE-IBS}, and \ac{PS-E2EID} according to the security definitions in Sec.~\ref{subsec:Securitydef1} and Sec.~\ref{subsec:Securitydef2}. Note that the \ac{OEEP-ABE} is reconfigured with EABEHP \cite{YuHsuLeeLee:20} and PEAPOD \cite{KapTsaSmi:07}, and the OE-IBS is also reconfigured with the IBS \cite{LjkBjZj:10}.
	\vspace{-3mm}
	\subsection{Security Analysis of OEEP-ABE and OE-IBS}\label{subsec:analysiscrypto}
	
	\begin{theorem}[Confidentiality of \ac{OEEP-ABE}]
		The proposed OEEP-ABE is with C-IND-CPA-RUCA if the decisional Diffie-Hellman~(DDH) assumption holds.
	\end{theorem}
	
	\textbf{Proof Sketch.}\footnote{The \ac{OEEP-ABE} cannot achieve collusion resistance, similar to EABEHP \cite{YuHsuLeeLee:20}. However, existing hardware security solutions (e.g., TPM and HSM) \cite{TPM,EVITA} are used in all entities of our system (e.g., \acp{ECU} and \acp{RSU}), so an adversary cannot obtain each device's secret. Hence, collusion attacks cannot be performed.} The proof for the confidentiality of the \ac{OEEP-ABE} consists of two parts: the confidentiality of ciphertexts against an outsider and that against the SAs. The outsider can obtain the ciphertext $C$ only, regardless of outsourcing. The most structure of $C$ is identical to that in EABEHP \cite{YuHsuLeeLee:20}, and the only difference is the existence of $(g^{dr})$ in $C$ in the \ac{OEEP-ABE}, which can be still secured by the DDH assumption.
	
	The SA has all the capabilities of an outsider and obtaining additional information by participating in the outsourcing procedure. The additional information is $v_{\tau}$ and $MO_{\tau}$ from \textbf{Out.Encrypt1}. To break the security of $C$, which was generated during the outsourced encryption, the preliminary ciphertext $pC$ needs to be computed. However, although the structure of $pC$ is similar to that of a general ciphertext, and $pC$ is a combination of $MO_{\tau}$ and $MO_{\tau '}$, which are respectively generated by the two SAs. Therefore, it is impossible to break the security of $pC$ under the DDH assumption. Furthermore, under the assumption of non-colluding SAs, the security can be maintained.

\begin{theorem}[Policy privacy of \ac{OEEP-ABE}]
	The proposed OEEP-ABE is with P-IND-CPA-RUCA if the DDH assumption holds.
\end{theorem}

	\textbf{Proof Sketch.} The proof for the policy privacy of the \ac{OEEP-ABE} also consists of two parts: the policy privacy against an outsider and that against the SAs. 
	For non-outsourced encryption in the \ac{OEEP-ABE}, a message $M$ is encrypted in the same way in \cite{YuHsuLeeLee:20} as follows: First, for each of $t_i$, $p_i$ is selected for $i \in I$ such that $\prod_{t_i = 1 \wedge i \in I} p_i = M$ for all $t_i = 1$, while $p_i = 1$ for $t_i = 0$, and $p_i=\alpha_i$ for $t_i = -1$, where $\alpha_i$ is a random number. Here, $t_i = 1, 0, -1$ if the attribute i is required, irrelevant, and forbidden, respectively. Then, each $p_i$ is encrypted using ElGamal encryption \cite{TsiYun:98}, so the knowledge on $p_i$ cannot be learned from the ciphertext $C$.
	
	The SAs need to obtain $pC$ to figure out the policy information. Similarly to the confidentiality of the \ac{OEEP-ABE}, it is impossible for the SAs to obtain $pC$ under the assumptions of DDH and non-colluding SAs, so the policy privacy of the OEEP-ABE cannot be broken by the outsider as well as the SAs.
	
\begin{theorem}[Unforgeability of OE-IBS]
		The proposed OE-IBS is with UF-IBS-CMA if the discrete logarithm~(DL) assumption holds.
\end{theorem}
	
	\textbf{Proof Sketch.} The structure of the OE-IBS is the same as that of the online/offline IBS in \cite{LjkBjZj:10}, except for the outsourced signing. Thus, we are required to prove the unforgeability against the SA participating in the outsourced signing. The SA can obtain $x_t$ and $Y'_t = Y^{x_t}$ from \textbf{Out.Sign1}, where $t$ is the message index. In other words, if the SA can obtain $z_t$, it can also obtain the user's secret key and forge the signature. However, the signature generator randomly selects $\mathcal{X}_t$ and $\mathcal{Y}_t$ such that $z_t = \mathcal{X}_t + \mathcal{Y}_t$, and produces a signature $\text{Sig}=(Y'_t, \mathcal{B}_j, \omega\mathcal{X}_t, g^{\frac{1}{\omega}}, \mathcal{Y}_t)$ using $\omega$ and $g^{\frac{1}{\omega}}$, generated by \textbf{Offline.Sign}. Here, $\omega$ is a secret value known only to the signature generator, so anyone (including the SA) cannot obtain $z_t$, even though $g^{z_t}$ can be obtained. Therefore, the unforgeability of the OE-IBS cannot be broken by the outsider as well as the SA.
\vspace{-3mm}
\subsection{Security Analysis of PS-E2EID}\label{sec:analysisprotocol}
This subsection proposes the security analysis of the proposed \ac{PS-E2EID} in Sec.~\ref{sec:protocol}. Here, we only provide the formal security analysis of the uplink ABDS protocol since the security of the other protocols including downlink ABDS and initial authentication and preliminary can also be proved similarly to that of the uplink ABDS protocol. We prove that the message authentication, attribute-based key exchange, policy privacy, and identity anonymity can be achieved.
	
\begin{theorem}[PS-E2EID Security] 
	The proposed PS-E2EID is said to be the anonymous attribute-based authenticated key exchange and data sharing protocol with hidden policy and traceability if $\text{H}$ is a pseudorandom function, $\text{E}_\text{S}$ is a pseudorandom permutation, the OEEP-ABE is C-IND-CPA-RUCA-seucre and P-IND-CPA-RUCA-secure attribute-based encryption, and the OE-IBS is UF-IBS-CMA-secure identity-based signature. Then, we have that
\end{theorem}
\vspace{-1mm}
\begin{align} \notag \label{eq:PS-E2EID}
	&\text{Adv}_{\mathcal{A}}^{\text{PS-E2EID}} \leq \frac{2}{3}\text{Adv}_{\text{H}}+\frac{2}{3}\text{Adv}_{\text{E}_\text{S}}+\frac{8}{3}{\text{Adv}_{\text{C-IND-CPA}}} \\ \notag
	&\quad \; +2{\text{Adv}_{\text{P-IND-CPA}}}+\frac{8}{3}\text{Adv}_{\text{DDH}}+\frac{4}{3}\text{Adv}_{\text{UF-IBS-CMA}}+\epsilon_{\text{MA}} \\ \notag
	&\epsilon_{\text{MA}} = \frac{4}{3}\text{Adv}_{\text{H}}\text{Adv}_{\text{E}_\text{S}}+\frac{4}{3}\text{Adv}_{\text{C-IND-CPA}}\text{Adv}_{\text{UF-IBS-CMA}}\\ 
	& \quad \qquad \qquad \qquad \qquad  +\frac{4}{3}\text{Adv}_{\text{DDH}}\text{Adv}_{\text{UF-IBS-CMA}}\approx 0.
\end{align}
where $\text{Adv}_{\mathcal{A}}^{\text{PS-E2EID}}$ is advantage that an attacker $\mathcal{A}$ breaks the security of the PS-E2EID. In addition, $\text{Adv}_{\text{H}}$, $\text{Adv}_{\text{E}_\text{S}}$, $\text{Adv}_{\text{DDH}}$, $\text{Adv}_{\text{C-IND-CPA}}$, $\text{Adv}_{\text{P-IND-CPA}}$, and $\text{Adv}_{\text{UF-IBS-CMA}}$ are advantages that break the security of the pseudorandom function, the pseudorandom permutation, the DDH assumption, the C-IND-CPA-RUCA of the OEEP-ABE, the P-IND-CPA-RUCA of the OEEP-ABE, and the UF-IBS-CMA of the OE-IBS, respectively.

\emph{Proof:} Let $\mathcal{A}$ be an adversary of breaking the security of message authentication, attribute-based key exchange, policy privacy, and identity anonymity of the proposed PS-E2EID. We perform security games for each four security features, mentioned Sec. II-C, to prove the security of the PS-E2EID. We then show that the advantages of $\mathcal{A}$ for the PS-E2EID can be negligible, depending on the advantages of $\mathcal{A}$ in each game.

\textit{Game $G_0$}: This is the real game, and $\mathcal{A}$ has access to the OEEP-ABE and the OE-IBS's master public key $(MPK,MSPK)$. In addition, $\mathcal{A}$ has the ability to query all oracles, specified in Sec.~\ref{subsec:Securitydef2}, and knows the structure of the PS-E2EID. Since the OEEP-ABE and the OE-IBS are proven to be secure with IND-CPA and UF-IBS-CMA in Sec.~\ref{subsec:analysiscrypto}, all the parameters, related to IND-CPA and UF-IBS-CMA, can be successfully simulated. Therefore, we have $\text{Adv}_{\mathcal{A}}^{\text{PS-E2EID}}=\text{Adv}_{\mathcal{A},0}^{\text{PS-E2EID}}$, where $\text{Adv}_{\mathcal{A},i}^{\text{PS-E2EID}}$ is the advantage of $\mathcal{A}$ in game $G_i$.

\textit{Game $G_1$} (Message Authentication):
In the game $G_1$, we describe the events as follows: $\text{E}_1$ is an event in which $\mathcal{A}$ impersonates $\text{ECU}_{nj}$ by sending the correct $C_M || \sigma_{M}$ to the $(\text{RSU}/\text{UE}/\text{OEM}/\text{OBU})_m$, and $\text{E}_2$ is an event in which $\mathcal{A}$ impersonates $\text{OBU}_{n}$ by sending the correct $\text{Token}^{\text{user}}_{n}$ to the $(\text{RSU}/\text{UE}/\text{OEM}/\text{OBU})_m$. 
We construct a simulator $S_1$ for the PS-E2EID that interacts with $\mathcal{A}$ as the security game, defined in Definition~\ref{def:5:Message-authentication}. In addition, $S_1$ is provided with the master public keys of the OEEP-ABE and the OE-IBS for the successful simulation. If $\text{E}_1$ happens, $S_1$ can exploit the ability of $\mathcal{A}$ to break some security. Specifically, the security analysis regarding $\text{E}_1$ can be divided into two cases. 

First, when $\text{E}_1$ happens, $S_1$ can break the security of the underlying pseudorandom function and the underlying pseudorandom permutation by exploiting the ability of $\mathcal{A}$. This event is denoted as $\text{E}_{11}$, and we have
\vspace{-1mm}
\begin{align} \notag \label{eq:gameg1-1}
	&\;\;\text{Adv}_{\text{H},\text{E}_\text{S}} \geq \{\text{Pr}[\text{S}_{\text{H},\text{E}_\text{S}},\text{E}_{11}]+\text{Pr}[\text{S}_{\text{H},\text{E}_\text{S}},\neg\text{E}_{11}]\}-\frac{1}{4}\\ \notag
	&=\hspace{-0.7mm} \{\text{Pr}[\text{S}_{\text{H},\text{E}_\text{S}}|\text{E}_{11}]\hspace{-0.7mm}\times\hspace{-0.7mm}\text{Pr}[\text{E}_{11}] \hspace{-0.7mm}+\hspace{-0.7mm}\text{Pr}[\text{S}_{\text{H},\text{E}_\text{S}}|\neg\text{E}_{11}]\hspace{-0.7mm}\times\hspace{-0.7mm}(1\hspace{-0.7mm}-\hspace{-0.7mm}\text{Pr}[\text{E}_{11}])\}\hspace{-0.7mm}-\hspace{-0.7mm}\frac{1}{4}\\ 
	& = \hspace{-0.7mm}	\{1\times\text{Adv}_{\text{E}_{11}}+\frac{1}{4}\times(1-\text{Adv}_{\text{E}_{11}})\}-\frac{1}{4} = \frac{3}{4}\text{Adv}_{\text{E}_{11}},
\end{align}
where $\text{S}_{\text{H},\text{E}_\text{S}}$ is the event that successfully distinguishing the pseudorandom function from a truly random function and the pseudorandom permutation from a truly random permutation simultaneously, $\text{Adv}_{\text{H},\text{E}_\text{S}}$ is the advantage that simultaneously breaks the security of the pseudorandom function and pseudorandom permutation, $\neg\text{E}_{11}$ is the complementary event of $\text{E}_{11}$, and $\text{Adv}_{\text{E}_{11}}$ is the advantage of $\text{E}_{11}$. In addition, $\text{Adv}_{\text{H},\text{E}_\text{S}}$ can be expressed using $\text{Adv}_{\text{H}}$ and $\text{Adv}_{\text{E}_\text{S}}$ as
\vspace{-1mm}
\begin{align} \notag \label{eq:gameg1-2}
	\text{Adv}_{\text{H},\text{E}_\text{S}} &= \text{Pr}[\text{S}_{\text{H},\text{E}_\text{S}}]-\frac{1}{4} = \text{Pr}[\text{S}_{\text{H}}]\text{Pr}[\text{S}_{\text{E}_\text{S}}]-\frac{1}{4}\\ \notag
	&=
	(\frac{1}{2}+\text{Adv}_{\text{H}})(\frac{1}{2}+\text{Adv}_{\text{E}_\text{S}})-\frac{1}{4}\\ 
	&= \frac{1}{2}(\text{Adv}_{\text{H}}+\text{Adv}_{\text{E}_\text{S}}) +\text{Adv}_{\text{H}}\text{Adv}_{\text{E}_\text{S}},
\end{align}
where $\text{S}_{\text{H}}$ is the event that successfully distinguishing the pseudorandom function from a truly random function, and $\text{S}_{\text{E}_\text{S}}$ is the event that successfully distinguishing the pseudorandom permutation from a truly random permutation. Thus, from \eqref{eq:gameg1-1} and \eqref{eq:gameg1-2}, we have $\text{Adv}_{\text{E}_{11}}\leq \frac{2}{3}(\text{Adv}_{\text{H}}+\text{Adv}_{\text{E}_\text{S}})+\frac{4}{3}\text{Adv}_{\text{H}}\text{Adv}_{\text{E}_\text{S}}$. 

Second, when $\text{E}_1$ happens, $S_1$ can also break the security of the underlying C-IND-CPA-RUCA-secure OEEP-ABE and the UF-IBS-CMA-secure OE-IBS by exploiting the ability of $\mathcal{A}$. This event is denoted as $\text{E}_{12}$. Thus, we have $\text{Adv}_{\text{E}_{12}}\leq \frac{2}{3}(\text{Adv}_{\text{C-IND-CPA}}+\text{Adv}_{\text{UF-IBS-CMA}})+\frac{4}{3}\text{Adv}_{\text{C-IND-CPA}}\text{Adv}_{\text{UF-IBS-CMA}}$. Based on the results of $\text{E}_{11}$ and $\text{E}_{12}$, we have
\vspace{-1mm}
\begin{align} \notag
	&\text{Adv}_{\text{E}_1} \leq \frac{2}{3}(\text{Adv}_{\text{H}}+\text{Adv}_{\text{E}_\text{S}}+\text{Adv}_{\text{C-IND-CPA}}+\text{Adv}_{\text{UF-IBS-CMA}}) \\ \notag
	& \qquad \qquad +\frac{4}{3}(\text{Adv}_{\text{H}}\text{Adv}_{\text{E}_\text{S}}+\text{Adv}_{\text{C-IND-CPA}}\text{Adv}_{\text{UF-IBS-CMA}}).
\end{align}
Then, if $\text{E}_2$ happens, $S_1$ can exploit the ability of $\mathcal{A}$ to break the security of the underlying DDH assumption and the UF-IBS-CMA-secure OE-IBS. Thus, we have $\text{Adv}_{\text{E}_2} \leq \frac{2}{3}(\text{Adv}_{\text{DDH}}+\text{Adv}_{\text{UF-IBS-CMA}}) +\frac{4}{3}\text{Adv}_{\text{DDH}}\text{Adv}_{\text{UF-IBS-CMA}}$.
Finally, we have
\vspace{-2mm}
\begin{align}\label{eq:gameg1} \notag
	&\text{Adv}_{\mathcal{A},0}^{\text{PS-E2EID}} \leq \text{Adv}_{\mathcal{A},1}^{\text{PS-E2EID}}+\frac{2}{3}\text{Adv}_{\text{H}}+\frac{2}{3}\text{Adv}_{\text{E}_\text{S}} \\ 
	&\hspace{-1.5mm}+\frac{2}{3}\text{Adv}_{\text{C-IND-CPA}}+\frac{4}{3}\text{Adv}_{\text{UF-IBS-CMA}}+\frac{2}{3}\text{Adv}_{\text{DDH}}+\epsilon_{\text{MA}},
\end{align}
where $\epsilon_{\text{MA}}$ is in \eqref{eq:PS-E2EID}.

\textit{Game $G_2$} (Attribute-based Key Exchange):
In the game $G_2$, we construct a simulator $S_2$ that interacts with $\mathcal{A}$ as the security game, defined in Definition~\ref{def:6:Attribute-based-Key-exchange}. Here, $S_2$ is provided with the master public key of the OEEP-ABE for the successful simulation. The $\mathcal{A}$ queries \textbf{Test} after interacting with $S_2$ as the security game, and $S_2$ responds to $\mathcal{A}$ with an attribute-based key $K$ or a random string according to a random bit. If $\mathcal{A}$ can successfully guess $K$, $S_2$ can break the security of the underlying C-IND-CPA-RUCA-secure OEEP-ABE by exploiting the ability of $\mathcal{A}$. Therefore, we have 
\vspace{-1mm}
\begin{align} \notag
	&\;\;\text{Adv}_{\text{C-IND-CPA}} \\ \notag
	& \qquad \geq \{\text{Pr}[\text{S}_{\text{C-IND-CPA}},\text{E}_\text{AKE}]+\text{Pr}[\text{S}_{\text{C-IND-CPA}},\neg\text{E}_\text{AKE}]\}-\frac{1}{2}\\ \notag
	&\qquad = \{\text{Pr}[\text{S}_{\text{C-IND-CPA}}|\text{E}_\text{AKE}]\times\text{Pr}[\text{E}_\text{AKE}] \\ \notag
	&\qquad \qquad 
	+\text{Pr}[\text{S}_{\text{C-IND-CPA}}|\neg\text{E}_\text{AKE}] \times(1-\text{Pr}[\text{E}_\text{AKE}])\}-\frac{1}{2}\\ \notag
	&\qquad =
	\{1\times\text{Adv}_{\text{E}_\text{AKE}}+\frac{1}{2}\times(1-\text{Adv}_{\text{E}_\text{AKE}})\}-\frac{1}{2}=\frac{\text{Adv}_{\text{E}_\text{AKE}}}{2},
\end{align}
where $\text{S}_{\text{C-IND-CPA}}$ is the event of winning C-IND-CPA-RUCA security game of OEEP-ABE,  $\text{E}_\text{AKE}$ is the event in which $\mathcal{A}$ distinguishes $K$ from the random string, and $\text{Adv}_{\text{E}_\text{AKE}}$ is the advantage of breaking attribute-based key exchange security. Therefore, we have $\text{Adv}_{\text{E}_\text{AKE}} \leq 2\text{Adv}_{\text{C-IND-CPA}}$. Thus, we have
\vspace{-1mm}
\begin{equation}\label{eq:gameg2}
	\text{Adv}_{\mathcal{A},1}^{\text{PS-E2EID}} \leq \text{Adv}_{\mathcal{A},2}^{\text{PS-E2EID}}+2\text{Adv}_{\text{C-IND-CPA}}.
\end{equation}

\textit{Game $G_3$} (Policy Privacy):
In the game $G_3$, we construct a simulator $S_3$ that interacts with $\mathcal{A}$ as the security game, defined in Definition~\ref{def:7:Policy-Privacy}. Here, $S_3$ is provided with the OEEP-ABE's master public key $MPK$ for the successful simulation. The $\mathcal{A}$ queries \textbf{TestPolicy} after interacting with $S_3$ as the security game, and $S_3$ responds to $\mathcal{A}$ with $P_0$ or $P_1$ according to a random bit. If $\mathcal{A}$ can successfully guess the correct policy, $S_3$ can break the security of the underlying P-IND-CPA-RUCA-secure OEEP-ABE by exploiting the ability of $\mathcal{A}$. Thus, we have
\vspace{-1mm}
\begin{align} \notag
	&\text{Adv}_{\text{P-IND-CPA}} \hspace{-0.7mm} \geq \hspace{-0.7mm}  \{\text{Pr}[\text{S}_{\text{P-IND-CPA}},\text{E}_\text{PP}]\hspace{-0.2mm}+\hspace{-0.2mm}\text{Pr}[\text{S}_{\text{P-IND-CPA}},\neg\text{E}_\text{PP}]\}\hspace{-0.5mm}-\hspace{-0.5mm}\frac{1}{2}\\ \notag
	&= \{\text{Pr}[\text{S}_{\text{P-IND-CPA}}|\text{E}_\text{PP}]\times\text{Pr}[\text{E}_\text{PP}] \\ \notag 
	&\quad
	+\text{Pr}[\text{S}_{\text{P-IND-CPA}}|\neg\text{E}_\text{PP}] \times(1-\text{Pr}[\text{E}_\text{PP}])\}-\frac{1}{2}\\ \notag
	&=
	\{1\times\text{Adv}_{\text{E}_\text{PP}}+\frac{1}{2}\times(1-\text{Adv}_{\text{E}_\text{PP}})\}-\frac{1}{2} = \frac{\text{Adv}_{\text{E}_\text{PP}}}{2},
\end{align}
where $\text{S}_{\text{P-IND-CPA}}$ is the event of winning P-IND-CPA-RUCA security game of OEEP-ABE, $\text{E}_\text{PP}$ is the event that $\mathcal{A}$ distinguishes the correct policy from the random string, and $\text{Adv}_{\text{E}_\text{PP}}$ is the advantage of breaking policy privacy. Thus, we have
\vspace{-1mm}
\begin{equation}\label{eq:gameg3}
	\text{Adv}_{\mathcal{A},2}^{\text{PS-E2EID}} \leq \text{Adv}_{\mathcal{A},3}^{\text{PS-E2EID}}+2{\text{Adv}_{\text{P-IND-CPA}}}.
\end{equation}
	\begin{table}[t!] \scriptsize
	\vspace{-5mm}
	\caption{Security comparison with related works}\vspace{-2mm}
	\begin{center}
		\renewcommand{\arraystretch}{1.4}
		\begin{tabular}{c|cccc}
			\specialrule{.1em}{.1em}{.1em}
			\multirow{2}{*}{} & \cite{WpCcmKs:20} & \cite{ShaLinLuZuo:16} & \cite{VijAzeKanDeb:16} & \cite{KanLiuYaoWanLi:16} \\ 
			& \cite{GouHarAliGue:21}  & \cite{FenYuAloAlaLvMum:20}  & \cite{TiaLiQuaChaBak:21} & Our work \\ \hline \hline
			\multirow{2}{*}{Message Authentication} & $\surd$ & $\surd$ & $\surd$ & $\surd$ \\
			& $\surd$ & $\times$ & $\surd$ & $\surd$         \\ \cline{1-1} \cdashline{2-5}
			\multirow{2}{*}{Data Confidentiality} & $\times$ & $\surd$ & $\surd$ & $\surd$ \\
			& $\surd$ & $\surd$ & $\surd$ & $\surd$             \\ \cline{1-1} \cdashline{2-5}
			\multirow{2}{*}{Identity Anonymity} & $\surd$ & $\times$ & $\times$ & $\surd$ \\
			& $\times$ & $\times$ & $\times$ & $\surd$                   \\ \cline{1-1} \cdashline{2-5}
			\multirow{2}{*}{Traceability} & $\surd$ & $\times$ & $\times$ & $\surd$ \\
			& $\times$ & $\times$ & $\times$ & $\surd$                    \\ \cline{1-1} \cdashline{2-5}
			\multirow{2}{*}{Fine-grained Access Control} & $\times$ & $\times$ & $\times$ & $\surd$ \\
			& $\surd$ & $\surd$ & $\surd$ & $\surd$                   \\ \cline{1-1} \cdashline{2-5}
			\multirow{2}{*}{Policy Privacy} & - & - & - & $\times$ \\
			& $\times$ & $\times$ & $\surd$ & $\surd$ \\ \cline{1-1} \cdashline{2-5}
			\multirow{2}{*}{\begin{tabular}[c]{@{}c@{}} E2E security for fine-grained access \\ control and authentication to messages \end{tabular}} & - & - & - & $\times$ \\
			& $\times$ & $\times$ & $\times$ & $\surd$ \\ \specialrule{.1em}{.1em}{.1em}
		\end{tabular}
	\end{center}
	\label{table:comparison}
	\vspace{-6mm}	
\end{table}	
\textit{Game $G_4$} (Identity Anonymity):
In the game $G_4$, we construct a simulator $S_4$ that interacts with $\mathcal{A}$ as the security game, defined in Definition~\ref{def:8:Identity-Anonymity}. The $\mathcal{A}$ queries \textbf{TestID} after interacting with $S_4$ as the security game, and $S_4$ responds to $\mathcal{A}$ with a target real identity or a randomly selected identity according to a random bit. If $\mathcal{A}$ can successfully guess the real identity, $\mathcal{A}$ has the advantage of breaking the underlying DDH assumption. Therefore, we have 
\vspace{-1mm}
\begin{equation} \notag
	\text{Adv}_{\text{E}_\text{PID}} \leq{} 2\text{Adv}_{\text{DDH}},
\end{equation}
where $\text{E}_\text{PID}$ is the event in which $\mathcal{A}$ distinguishes the real target identity from the randomly selected identity, and $\text{Adv}_{\text{E}_\text{PID}}$ is the advantage of breaking identity anonymity. Thus, we have
\vspace{-1mm}
\begin{equation}\label{eq:gameg4}
	\text{Adv}_{\mathcal{A},3}^{\text{PS-E2EID}} \leq \text{Adv}_{\mathcal{A},4}^{\text{PS-E2EID}}+2\text{Adv}_{\text{DDH}}.
\end{equation}

There are no additional advantages other than those analyzed in the games $G_0$ to $G_4$. Thus, by equations \eqref{eq:gameg1}, \eqref{eq:gameg2}, \eqref{eq:gameg3}, and \eqref{eq:gameg4}, and $\text{Adv}_{\mathcal{A}}^{\text{PS-E2EID}}=\text{Adv}_{\mathcal{A},0}^{\text{PS-E2EID}}$, we conclude that the advantages of $\mathcal{A}$ to the PS-E2EID are given by \eqref{eq:PS-E2EID}.

Finally, the overall security comparison between the \ac{PS-E2EID} and existing related works is shown in Table \ref{table:comparison}. From Table \ref{table:comparison}, we can see that our proposed protocol can ensure more security features than other existing works.
\vspace{-2mm}
\section{Performance Analysis}\label{sec:performanceanalysis}
We prove the practicality of \ac{PS-E2EID} by evaluating the executions time of the cryptographic algorithms at system entities and that of the proposed protocols. The testbed consists of a desktop, a laptop, and a micro control unit. The desktop is with Intel\textregistered~Core\texttrademark~i5-11500 @ 2.7 GHz 6 cores processor and 16 GB RAM. The laptop is with Intel\textregistered~Core\texttrademark~i3-1115G4 @ 3.0 GHz 2 cores processor and 8 GB RAM. As a micro control unit, TI TMS320C28346 is with 300 MHz clock speed. Besides, we adopt the CAN-FD standard \cite{Har:12} for our in-vehicle networks, which is implemented by CANoe \cite{Vector}.
\vspace{-2mm}

\subsection{Cryptographic Algorithm Execution time}
			\begin{table}[t!] \scriptsize
	\vspace{-5mm}
	\caption{Execution time of \ac{OEEP-ABE} and \ac{OE-IBS}}\vspace{-2mm}
	\begin{center}
		\renewcommand{\arraystretch}{1.4}
		\begin{tabular}{c|c|c|c|c|c}
			\specialrule{.1em}{.1em}{.1em}
			\multicolumn{2}{c|}{OEEP-ABE} & \multicolumn{4}{c}{\textbf{Algorithm execution time [ms]}} \\  \hline
			\multicolumn{2}{c|}{{\textbf{Number of system attributes}}} & 4 & 8 & 16 & 32 \\ \hline
			Encrypt & ADAS, RSU, UE, OEM & 0.5 & 0.8 & 1.4 & 2.7 \\  \hline
			Out.Encrypt1 & SA(OBU)  & 0.8 & 1.4 & 2.6 & 4.9 \\  \hline
			Out.Encrypt2 & ECU  & 9.7 & 16.2 & 29.2 & 55.2\\  \hline 
			Select.Policy & ECU & 0.7 & 1.4 & 2.7 & 5.4 \\ \hline \hline		
			\multicolumn{2}{c|}{\textbf{Number of receiver attributes}} & 4 & 8 & 16 & 32 \\  \hline
			Decrypt	& ECU & 292.4 & 298.9 & 311.9 & 337.9 \\  \hline 
			Decrypt	& ADAS, RSU, UE, OEM & 0.16 & 0.17 & 0.2 & 0.24 \\  \hline \hline
			\multicolumn{2}{c|}{OE-IBS} & \multicolumn{4}{c}{\textbf{Algorithm execution time [ms]}} \\  \hline
			Offline.Sign & ECU & \multicolumn{4}{c}{284.3} \\  \hline
			Sign & ADAS, RSU, UE, OEM & \multicolumn{4}{c}{0.08} \\  \hline
			Out.Sign1 & SA(OBU) & \multicolumn{4}{c}{0.14} \\  \hline
			Out.Sign2 & ECU & \multicolumn{4}{c}{1.62} \\  \hline
			Verify & ECU & \multicolumn{4}{c}{430} \\  \hline
			Verify & ADAS, RSU, UE, OEM & \multicolumn{4}{c}{0.31} \\  \hline
		\end{tabular}
	\end{center}
	\label{table:OEEPABE}
	\vspace{-6mm}
\end{table}
For the operations of variables, encryption, and hash functions, we exploit BigInteger, SHA, AES, and modPow functions in \ac{JCA}/\ac{JCE}, and then evaluate the execution times of the cryptographic algorithms. We use the desktop as \ac{UDM}/\ac{ARPF}, \ac{ADAS}, \acp{RSU}, \acp{UE}, and \acp{OEM}, the laptop as the \ac{SA}, and the micro control units as an \acp{ECU}.
We implement the OEEP-ABE and the OE-IBS, and especially, the OEEP-ABE is implemented similarly to ElGamal encryption \cite{TsiYun:98}, whose algorithm execution times are demonstrated in Table.~\ref{table:OEEPABE}.\footnote{Note that the \ac{ECU} cannot conduct exponential and multiplicative operations in $\mathbb{G}$ due to its hardware limitation. Therefore, we first measure their execution times on the laptop, and then scale up them based on the increasing ratio of the other operations' execution times (e.g., SHA-256 and AES128) in \acp{ECU}. We expect that \acp{ECU} will have sufficient capabilities to perform advanced cryptographic operations in near future.} Note that each captured execution time is averaged over 10,000 iterated test runs.

\vspace{-3mm}
\subsection{Security Protocol Evaluation}\label{protocolevaluation}

\begin{figure*}[t!]
	\vspace{-8mm}
	\centering
	\begin{subfigure}[b]{0.325\textwidth}
		\centering
		\captionsetup{justification=centering}
		\includegraphics[width=\textwidth]{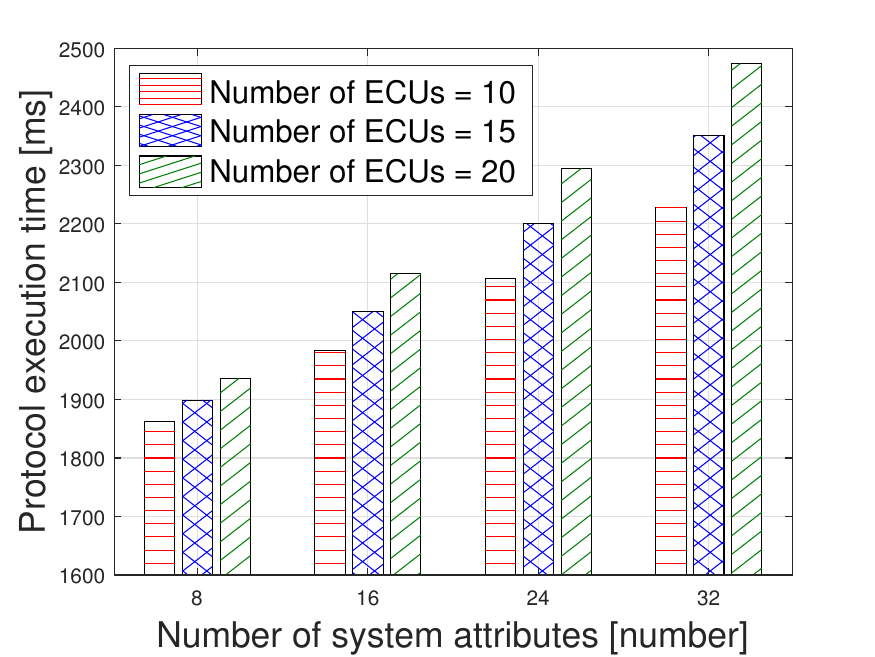}
		\caption{Initial authentication and preliminary protocol}
		\label{sublabel1}
	\end{subfigure}
	\hfill
	\begin{subfigure}[b]{0.325\textwidth}
		\centering
		\captionsetup{justification=centering}
		\includegraphics[width=\textwidth]{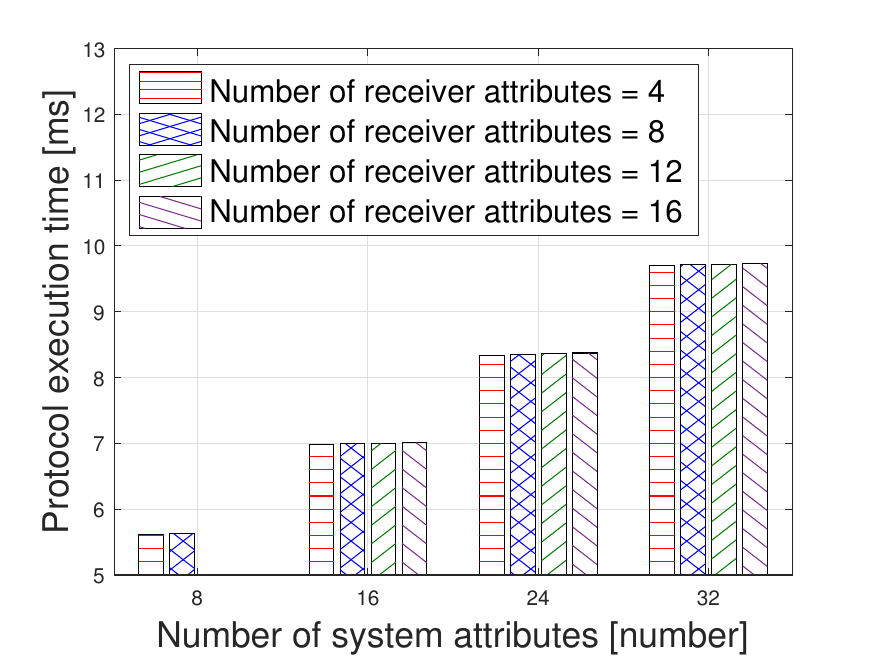}
		\caption{Uplink ABDS protocol \newline}
		\label{sublabel2}
	\end{subfigure}
	\hfill
	\begin{subfigure}[b]{0.325\textwidth}
		\centering
		\captionsetup{justification=centering}
		\includegraphics[width=\textwidth]{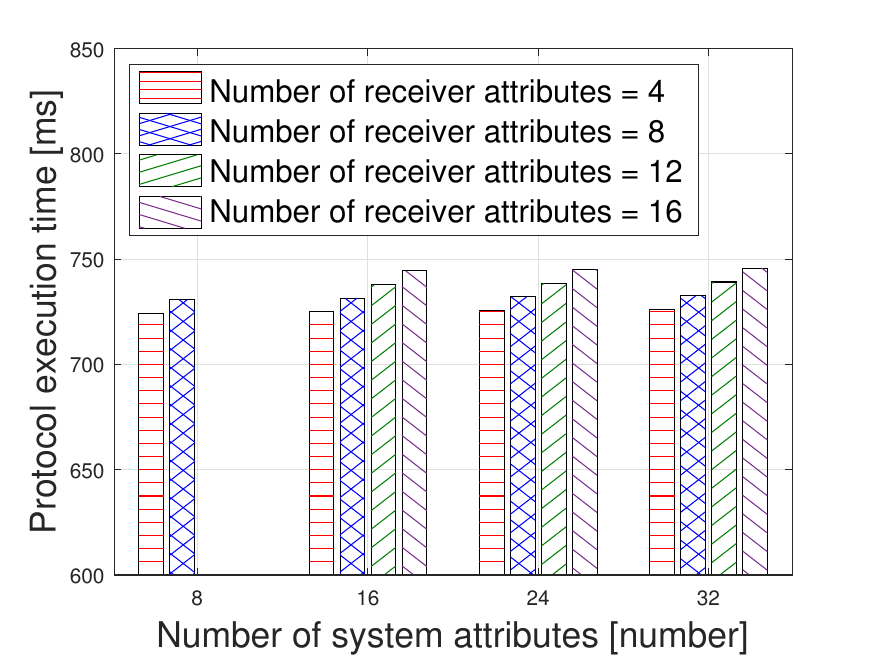}
		\caption{Downlink ABDS protocol \newline}
		\label{sublabel3}
	\end{subfigure}
	\vspace{-3mm}
	\caption{Execution time of PS-E2EID}
	\vspace{-5mm}
	\label{label}
\end{figure*}	

	We use 16 Bytes truncated MAC for in-vehicle communications \cite{WooJoKimLee:16,YuHsuLeeLee:20}, so the input size for SHA256 and AES128 is 48 Bytes. In addition, SHA256 and AES256 are used for hashing and encryption in \ac{V2X} communications. Figure~\ref{sublabel1} shows the execution times of the initial authentication and preliminary protocol for different numbers of system attributes and ECUs in the CAV. We consider the communication time as 1 ms in 5G-based C-V2X networks \cite{SSHus:21}.
	Specifically, as the numbers of \acp{ECU} and messages types (i.e., policies) increase, the number of \textbf{Out.Encrypt1} execution linearly increases. As the number of messages types increases, the numbers of \textbf{Out.Encrypt2} and \textbf{Offline.Sign} execution also linearly increase.
	For simplicity, we assume that each \ac{ECU} can have 5 message types and the data phase bit rate in in-vehicle networks is 8 Mbit/s.
	As a result, the initial authentication and preliminary protocol require about 3.2 seconds for execution when 16 system attributes and 100 \acp{ECU} exist in the CAV. This result shows that the initial authentication and preliminary protocol can be finished within a reasonable time without hindering vehicle functions and user convenience as this protocol can be conducted in the \textit{starting state} of the CAV.

	Figures~\ref{sublabel2} and ~\ref{sublabel3} demonstrate the execution times of the uplink and downlink ABDS protocols, respectively. In the uplink ABDS protocol, \acp{ECU} can quickly encrypt and sign data using the preliminary ciphertext and signature materials, generated in the initial authentication and preliminary phase. 

	As shown in Figs.~\ref{sublabel2} and ~\ref{sublabel3}, the execution time of uplink and downlink ABDS protocol are more affected dominantly by the number of system attributes and receiver attributes, respectively. Specifically, for the uplink ABDS protocol, the outsourced encryption is conducted for resource-constrained ECUs, and the lightweight decryption of the OEEP-ABE is conducted for high-performance nodes (e.g., RSUs, UEs, OEMs, and OBUs). However, even though the ECUs use the outsourced encryption, the high-performance nodes can perform the lightweight decryption faster, so the execution time for encryption is longer than that of decryption in the uplink ABDS protocol. Moreover, the computation cost of encryption increases as the number of system attributes increases, so the execution time increases as well. Hence, as shown in Fig.~\ref{sublabel2}, the execution time increases more with the number of system attributes rather than the number of receiver attributes. On the other hand, for the downlink ABDS protocol, the encryption is conducted for high-performance nodes, and the lightweight decryption of the OEEP-ABE is conducted for resource-constrained ECUs. However, even though the ECUs perform lightweight decryption, the high-performance nodes can perform the encryption faster, so the execution time for decryption is longer than that of encryption in the downlink ABDS protocol. Moreover, the computation cost of decryption increases as the number of receiver attributes increases, so the execution times increase as well. Hence, as shown in Fig.~\ref{sublabel3}, the execution time increases more with the number of receiver attributes rather than the number of system attributes.
	
	Additionally, the execution time of message signing is relatively small in the uplink and downlink ABDS protocols as shown in Table~\ref{table:OEEPABE}. This is because the message signing requires lower computation cost, compared to the message encryption, and especially, the execution time can be even smaller when ECUs use the outsourced signing in the OE-IBS. Note that the OE-IBS supports batch verification, so the verification time can be further smaller than other works.
	
	Finally, we evaluate the execution times of the uplink and downlink ABDS protocols at two types of in-vehicle end-devices with the sufficient numbers of system attributes and receiver attributes to classify entities in our system (i.e., 16 system attributes and 8 receiver attributes). Specifically, for the uplink and downlink ABDS protocols, the execution times are about 7.15 ms and 731 ms at an ECU\footnote{In the case of the downlink ABDS protocol, the ECU generally receives data from outside nodes for non-time-critical purposes such as firmware updates or speed control for the road speed limit. Hence, a certain amount of latency (e.g., less than 1 second) can be allowed.}, respectively, and 4.85 ms and 6.09 ms at the ADAS, respectively. Therefore, the proposed PS-E2EID can be practically used to ensure the \emph{E2E security} for vehicular communications.
	\vspace{-3mm}
	\subsection{Performance Comparison with Related Works}
	
	\begin{table*}[t!]\scriptsize
		\vspace{-5mm}
		\caption{Performance comparison with related works}
		\vspace{-2mm}
		\begin{center}
			\renewcommand{\arraystretch}{1.4}
			\begin{tabular}{|c||c||c||c||c|}
				\hline
				& Our work & Wang \emph{et al.} \cite{WpCcmKs:20} & Shao \emph{et al.} \cite{ShaLinLuZuo:16} & Tian \emph{et al.} \cite{TiaLiQuaChaBak:21} \\ \hline
				Sender's computation cost & $(N_s +3) T_{\text{M}}+ 3T_{\text{H}} + 4T_{\text{E}_\text{S}}$ & $T_{\text{EXP}}+T_{\text{RSA-V}}+T_{\text{RSA-D}}$ & $14T_{\text{EXP}}+3T_{\text{P}}$ & $|K|T_{\text{M}}$\\ \hline
				Computation time ($N_s, |K| = 16$) & 4.22 ms (Sender : ECU) & 7.55 ms (Sender : OBU) & 2.26 ms (Sender : OBU) & 0.07 ms (Sender : OBU) \\ \hline
				Receiver's computation cost & $11T_{\text{EXP}} + (N_r + 8)T_{\text{M}}+ 6T_{\text{H}}+ 2T_{\text{E}_\text{S}}$ & $T_{\text{RSA-V}}+T_{\text{RSA-E}}+T_{\text{RSA-D}}$ & $6T_{\text{EXP}}+13T_{\text{P}}$ & $2T_{\text{EXP}}+T_{\text{M}}$\\ \hline
				Computation time ($N_r = 8$) & 0.87 ms & 4.02 ms & 1.03 ms & 0.15 ms \\ \hline 
				\multicolumn{5}{|l|}{\begin{tabular}[l]{@{}l@{}} $T_{\text{EXP}}:$ computation cost of exponential operation, $T_{\text{M}}:$ computation cost of multiplicative operations, $T_{\text{H}}:$ computation cost of hash function \\ $T_{\text{E}_\text{S}}:$ computation cost of symmetric encryption, $N_s:$ number of system attributes in \ac{OEEP-ABE}, $N_r:$ number of receiver attributes in \ac{OEEP-ABE} \\ $N_m:$ number of messages to be encrypted in the ABDS protocol, $|K|:$ number of attributes for encryption in \cite{TiaLiQuaChaBak:21}, $T_{\text{P}}:$ computation cost of pairing operation \\ $T_{\text{RSA-E}}:$ computation cost of RSA encryption, $T_{\text{RSA-D}}:$ computation cost of RSA decryption, $T_{\text{RSA-V}}:$ computation cost of RSA signature verification \end{tabular}} \\ \hline
			\end{tabular}
		\end{center}
		\label{table:compperformance}
		\vspace{-5mm}
	\end{table*}	
	In this subsection, we compare the uplink ABDS protocol in PS-E2EID and the existing solutions in \cite{WpCcmKs:20,ShaLinLuZuo:16,TiaLiQuaChaBak:21}.\footnote{As explained in Sec.~\ref{sec:analysisprotocol}, the uplink ABDS protocol is the most significant part of our work. Therefore, we select the uplink ABDS protocol for the performance comparison.} Note that we adopt V2X communication-related works for the comparison since our work is the \textit{pioneering work} to jointly consider vehicular communications, i.e., in-vehicle and V2X communications together. Table.~\ref{table:compperformance} shows the computation costs and times of the uplink ABDS protocol, compared to those in \cite{WpCcmKs:20,ShaLinLuZuo:16,TiaLiQuaChaBak:21} at sender and receiver sides, respectively. Note that the sender in our work is an \ac{ECU}, and those in other works are the OBUs (as the existing work do not consider E2E security to in-vehicle end-devices).
	
	The our work uses the outsourced encryption and signing, so the sender's computation cost is lower than those in \cite{WpCcmKs:20, ShaLinLuZuo:16}. In addition, in case of the encryption, the outsourced encryption in \cite{TiaLiQuaChaBak:21} has the same cost $N_s T_{\text{M}}$ as our work. Here, the sender's computation time of our work can be higher than those in \cite{ShaLinLuZuo:16, TiaLiQuaChaBak:21}, since our work considers ECU as a sender, different from OBUs in other works.
	
On the other hand, in the aspect of the receiver's cost, the cost of our work is lower than that of the works \cite{WpCcmKs:20, ShaLinLuZuo:16}, but is higher than that in \cite{TiaLiQuaChaBak:21}. Note that the receiver's computation cost in \cite{TiaLiQuaChaBak:21} is for the decryption only, while that in our work is for a protocol to achieve additional security features, including the decryption. Compareing the decryption cost at the receiver only, the cost of our work is $2T_{\text{EXP}}+(N_r+1)T_{\text{M}}$ and that in \cite{TiaLiQuaChaBak:21} is  $2T_{\text{EXP}}+T_{\text{M}}$, which makes $N_r T_M$ difference. Here, $T_M$ is generally quite small as it is the computation time for multiplicative operations, so the difference $N_r T_M$ is not significant. Furthermore, different from our work, as the outsource decryption is used in \cite{TiaLiQuaChaBak:21}, it requires not only the computation at the receiver, but also that at the SA. The computation cost is large at the SA, and it results in long latency during the data sharing in the \textit{driving state}.

In the aspect of the encryption, our work uses the outsource encryption while \cite{TiaLiQuaChaBak:21} does not. However, as the outsourcing encryption is performed off-line ahead, the computation time for the encryption in the \textit{driving state} is significantly smaller than that in \cite{TiaLiQuaChaBak:21}. Hence, we can see that our proposed protocol can be more suitable in the aspect of the user convenience in CAVs, compared to other works including \cite{TiaLiQuaChaBak:21}, and it provides additional security features, including the E2E security to in-vehicle end-devices and identity anonymity, with sufficiently low computation cost for vehicular communications.

	\section{Conclusions}
	This work proposes the practical and secure vehicular communication protocol for E2E security to in-vehicle end-devices (PS-E2EID) with attribute-based access control and policy privacy. 
	Specifically, in PS-E2EID, we propose the OEEP-ABE to achieve attribute-based access control and policy privacy with low computation cost, and the OE-IBS to reduce the computation cost in message signing at in-vehicle end-devices.
	The security of the PS-E2EID is then proved based on the security of the pseudorandom function, pseudorandom permutation, C-IND-CPA-RUCA and P-IND-CPA-RUCA of the OEEP-ABE, and UF-IBS-CMA of the OE-IBS.
	Besides, we then evaluate the performance of the PS-E2EID with different numbers of sending ECUs, system attributes, and receiver attributes. The experimental results show that the PS-E2EID can achieve a high level of security in a reasonable time for the CAV communications, where in-vehicle end-devices including resource-constrained ECUs work as senders and receivers. 
	Finally, through the performance evaluation, we show that the PS-E2EID can be practically adopted to ensure the E2E security to in-vehicle end-devices.

	\bibliographystyle{bib/IEEEtran}
	\bibliography{bib/StringDefinitions,bib/IEEEabrv,bib/mybib}
	
\end{document}